\documentclass[%
 reprint,
superscriptaddress,
 amsmath,amssymb,
 aps,
pra,
]{revtex4-2}

\usepackage{graphicx}% Include figure files
\usepackage{dcolumn}% Align table columns on decimal point
\usepackage{bm}% bold math
\usepackage{hyperref}% add hypertext capabilities
\usepackage{here}
\def\Kopt{\bm{K}_\mathrm{opt}}
\def\Kmech{\bm{K}_\mathrm{mech}}

\usepackage{color}

\begin{document}

\preprint{APS/123-QED}

\title{Angular trapping of a linear-cavity mirror with an optical torsional spring}
%Geometrical optical trapping of a linear-cavity mirror in the rotational degree of freedom

\author{Takuya~Kawasaki}
  \email{takuya.kawasaki@phys.s.u-tokyo.ac.jp}
 \affiliation{Department of Physics, University of Tokyo, Bunkyo, Tokyo 113-0033, Japan}%Lines break automatically or can be forced with \\
\author{Kentaro~Komori}%
 \affiliation{Department of Physics, University of Tokyo, Bunkyo, Tokyo 113-0033, Japan}
 \affiliation{Institute of Space and Astronautical Science, Japan Aerospace Exploration Agency, Sagamihara, Kanagawa 252-5210, Japan}
\author{Hiroki~Fujimoto}%
 \affiliation{Department of Physics, University of Tokyo, Bunkyo, Tokyo 113-0033, Japan}
\author{Yuta~Michimura}
 \affiliation{Department of Physics, University of Tokyo, Bunkyo, Tokyo 113-0033, Japan}
\author{Masaki~Ando}
 \affiliation{Department of Physics, University of Tokyo, Bunkyo, Tokyo 113-0033, Japan}
 \affiliation{Research Center for the Early Universe (RESCEU), University of Tokyo, Bunkyo, Tokyo 113-0033, Japan}

\date{\today}

\begin{abstract}
Optomechanical systems have attracted intensive attention in various physical experiments.
With an optomechanical system, the displacement of or the force acting on a mechanical oscillator can be precisely measured by utilizing optical interferometry.
As a mechanical oscillator, a suspended mirror is often used in over a milligram scale optomechanical systems.
% Suspension with a thin wire is resistant to environmental noises.
However, the tiny suspended mirror in a linear cavity can be unstable in its yaw rotational degree of freedom due to optical radiation pressure.
This instability curbs the optical power that the cavity can accumulate in it, and imposes a limitation on the sensitivity.
Here, we show that the optical radiation pressure can be used to trap the rotational motion of the suspended mirror without additional active feedback control when the $g$ factors of the cavity are negative and one mirror is much heavier than the other one.
Furthermore, we demonstrate experimentally the validity of the trapping.
We measured the rotational stiffness of a suspended tiny mirror with various intracavity power.
The result indicates that the radiation pressure of the laser beam inside the cavity actually works as a positive restoring torque.
Moreover, we discuss the feasibility of observing quantum radiation pressure fluctuation with our experimental setup as an application of our trapping configuration.

% \begin{description}
% Secondary publications and information retrieval purposes.
% You may use the \texttt{description} environment to structure your abstract;
% use the optional argument of the \verb+\item+ command to give the category of each item. 
% \end{description}
\end{abstract}

\maketitle

%\tableofcontents

\section{\label{sec:introduction}Introduction}
% Precise measurement by optomechanical systems.
Optomechanical systems are widely used in experiments of precise measurements~\cite{Aspelmeyer:2013lha}.
In particular, optomechanical systems consisting of massive oscillators are suitable for testing macroscopic quantum mechanics~\cite{Chen:2013sgs}, investing Newtonian interaction of quantum objects~\cite{Miao:2019pxw, Kafri:2014zsa}, measuring gravitational force of milligram masses~\cite{Schmole:2016mde}, and gravitational wave detection~\cite{LIGOScientific:2016aoc}.
An optomechanical system consists of mechanical oscillators coupled with optical fields.
The displacement of or the force acting on the oscillator can be measured precisely by using optical interferometry.
Massive oscillators not only make the optomechanical systems resistant to noise sources~\cite{LIGOScientific:2016aoc}, but also allow for the exploration of the unique physics~\cite{Chen:2013sgs,Miao:2019pxw,Kafri:2014zsa,Schmole:2016mde}.
Ultimately, macroscopic quantum optomechanical systems are expected to elucidate the quantum nature of gravity~\cite{Marletto:2017kzi, Bose:2017nin, Belenchia:2018szb}.

% Therefore, even if the probe object is massive, its displacement can be measured with an accuracy of the wavelength scale of the light.

% Uses/Advantages of optomechanical systems as a sensor.
% - pure sensor -- GW detector, DM search
% - Macroscopic detector > MQM, Quantum gravity
% Recently, methods to detect ultralight dark matter were proposed as a force sensor.

% Previous research は？
% over mg optomechanical systems > pendulum
% Issue of pendulum - Sidles-Sigg instability
Suspended pendulums are often used as mechanical oscillators in optomechanical systems over a milligram scale~\cite{Pontin:2016nem, Matsumoto:2013sua, Matsumoto:2014fda, Komori:2019zlg, Sakata:2010zz, Matsumoto:2018via, Corbitt:2007spn, Neben:2012edt, LIGOScientific:2009mif, LIGOScientific:2020luc}, while membranes and cantilevers are used in many experiments of smaller mass scales~\cite{Chan:2011ivv, Teufel:2011smx, Peterson:2016ayo}.
Suspended pendulums are advantageous in that they can be isolated from the environment.
In other words, pendulums are resistant to seismic noise and thermal noise.
Furthermore, a pendulum can be regarded as a free mass in the broad frequency range over the resonant frequency.
In general, pendulums have a wide range of sensitivity for this reason. 
Recently, detecting schemes for ultralight dark matter using optomechanical oscillators were proposed~\cite{Graham:2015ifn, Pierce:2018xmy, Carney:2019cio}, and the wide searchable range was crucial for the dark matter search because the mass of dark matter was scarcely known.

\begin{figure}
    \centering
    \includegraphics[width=0.7\columnwidth,clip]{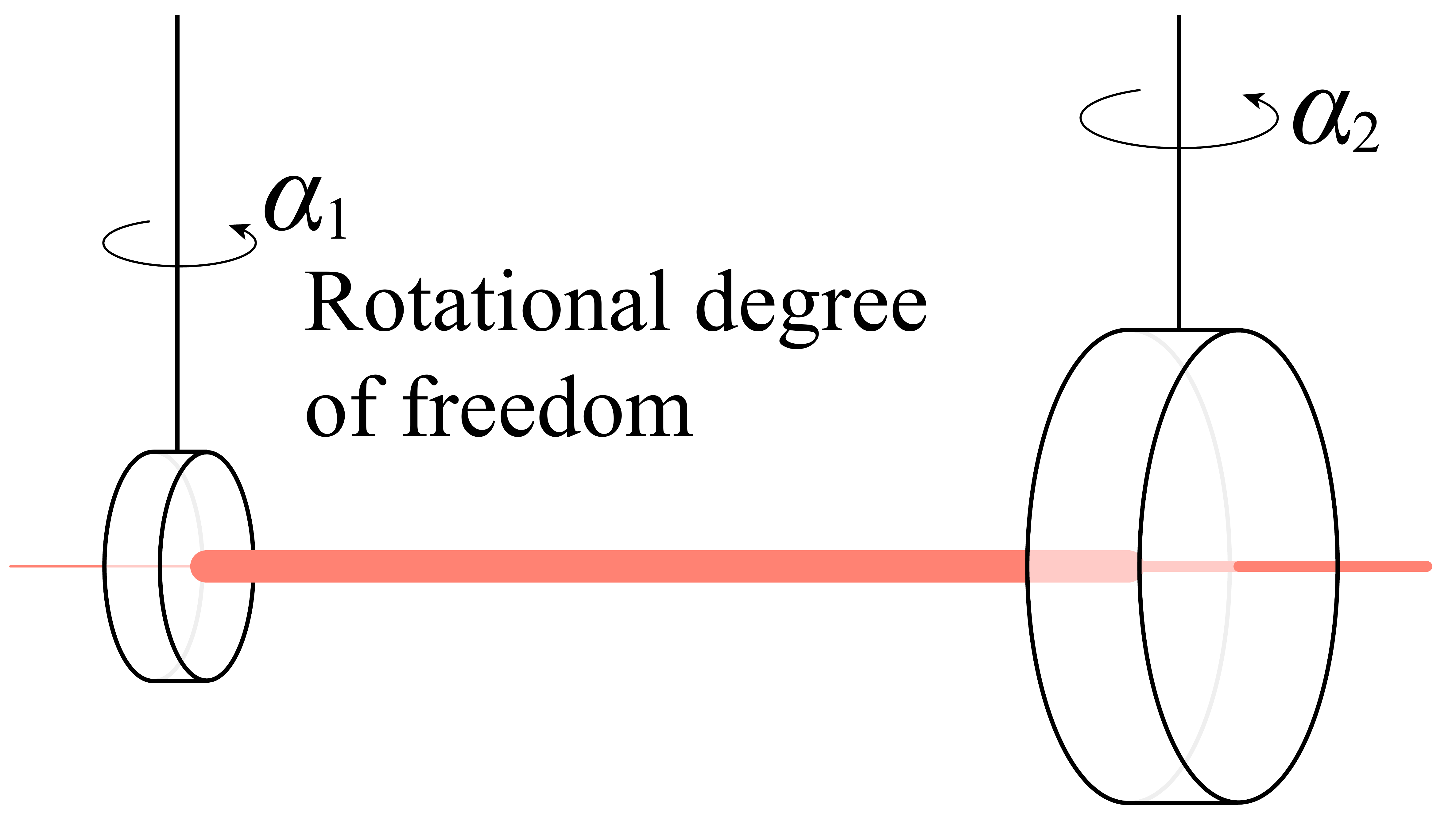}
    \caption{Schematic illustration of a suspended linear cavity. The rotational degrees of freedom we focus on are defined along the vertical axis.
    The input mirror is much heavier than the end mirror for our configuration.}
    \label{fig:rotation}
\end{figure}
However, the issue of suspended pendulums as mechanical oscillators is their instability in the rotational degree of freedom, the so-called Sidles-Sigg instability~\cite{Sidles:2006vzf}.
In a linear cavity, the yaw rotational motion of a suspended mirror that is indicated in Fig.~\ref{fig:rotation} changes the position of the beam spot of the laser beam.
Due to the change in the position of the beam spot, the radiation pressure in the cavity can behave as an anti-restoring force and destabilize the cavity.

In the case that the mechanical restoring torque is dominant in the rotational degree of freedom, the Sidles-Sigg instability does not matter because the radiation pressure torque would be too weak to make a pendulum unstable.
Thus, several experiments used multiple wires to suspend a mirror to stiffen the pendulum in the rotational degree of freedom~\cite{Corbitt:2007spn, Neben:2012edt, Kelley:2015axa}.
However, the increase of wires induces a stronger coupling to a thermal bath.
As a result, the thermal noise due to the suspension gets larger.
% In addition, suspension with multiple wires is technically difficult to suspend a pendulum symmetrically.

One other way to deal with Sidles-Sigg instability is introducing feedback control in the rotational motion~\cite{Barsotti:2010zz}.
If an oscillator has actuators applying torque on it, the unstable rotational motion can be suppressed.
Therefore, modern gravitational wave detectors use active feedback controls for the angular motion~\cite{Liu:2018dfs}.
However, active feedback control systems were not always available, especially for macroscopic quantum experiments and quantum measurements.
This was because milligram or gram scale oscillators were often too small to attach actuators to~\cite{Sakata:2010zz}.
In such cases, a more complicated feedback system was required to actuate a tiny oscillator remotely~\cite{Enomoto:2016lee, Nagano:2016wcg}.

As a different approach, some experiments used triangular cavities to avoid the Sidles-Sigg instability~\cite{Matsumoto:2013sua, Matsumoto:2014fda, Komori:2019zlg}.
When the number of mirrors consisting in the cavity is odd, the radiation pressure behaves as a positive restoring torque.
Thus, the Sidles-Sigg instability can be avoided.
However, additional mirrors are noise sources~\cite{Komori:2019zlg}.
% Actually, the sensitivity in the previous experiment using triangular cavity was limited by thermal noises of the additional mirror in the high frequency range~\cite{Komori:2019zlg}.
For the best sensitivity, a cavity with only two mirrors (a linear cavity) is favorable.
Therefore, a stable configuration of a linear cavity is desired.

% in this paper.
In this paper, we propose and experimentally validate a stable trapping configuration for a suspended mirror in a linear cavity.
We utilize radiation pressure inside the cavity as the restoring torque.
Thus, our system can trap a rotational motion of the suspended mirror without additional active feedback controls.
Therefore, our trapping scheme is applicable to tiny systems that cannot have actuators attached to them, and it is free from the feedback-control noises.
In our system, we operate the cavity in the negative-$g$ regime, and the input mirror is much heavier than the end test mass.
The combined characteristics of the negative-$g$ regime and unbalanced masses produce a stable rotational trapping.
Furthermore, we validate the configuration experimentally with an 8 mg mirror.
In addition, we discuss the feasibility to observe the quantum radiation pressure fluctuation of a milligram scale optomechanical system for testing macroscopic quantum mechanics by using our trapping configuration.

\section{\label{sec:theory}Theoretical description}
We analyze the rotational motion of suspended mirrors in a linear cavity.
The following calculation shows the suspended mirrors are trapped with the positive radiation pressure torque under the condition that the cavity is in the negative-$g$ regime and one mirror is much heavier than the other one.
As shown in Fig.~\ref{fig:rotation}, we define the angles of two mirrors as $\alpha_i$ and the torques exerted to them as $T_i$ ($i = 1,~2)$.
The equation of motion of the rotational modes of the two mirrors is given by
\begin{align}
    (\Kopt + \Kmech - \bm{I}\omega^2)
    \begin{pmatrix}
    \alpha_1 \\ \alpha_2
    \end{pmatrix}
    =
    \begin{pmatrix}
    T_1 \\ T_2
    \end{pmatrix}
    ,
    \label{eq:motion}
\end{align}
where
\begin{align}
    \Kmech =
    \begin{pmatrix}
    K_1 & 0 \\
    0   & K_2
    \end{pmatrix}
    ,~~
    \bm{I} =
    \begin{pmatrix}
    I_1 & 0 \\
    0   & I_2
    \end{pmatrix}
\end{align}
are the matrices of the mechanical restoring torques and the moment of inertia of each mirror.
The optical torsional stiffness matrix is represented as~\cite{Sidles:2006vzf}
\begin{align}
    \Kopt = \frac{2P}{c(R_1 + R_2 - L)}
    \begin{pmatrix}
    R_1(L - R_2) & R_1R_2 \\
    R_1R_2       & R_2(L - R_1)
    \end{pmatrix},
    \label{eq:stiffness}
\end{align}
where $P$, $c$, $R_1$, and $L$ are the intracavity power, the speed of light, the radii of curvature, and the cavity length, respectively.
The optical torque is generated by the change of the beam spot on the mirror due to the movement of the cavity axis as the mirror rotates.
Equation~(\ref{eq:motion}) can be rewritten as
\begin{align}
    \begin{pmatrix}
    K_1 - \beta g_2 - I_1\omega^2 & \beta \\
    \beta                         & K_2 - \beta g_1 -I_2\omega^2
    \end{pmatrix}
    \begin{pmatrix}
    \alpha_1 \\ \alpha_2
    \end{pmatrix}
    =
    \begin{pmatrix}
    T_1 \\ T_2
    \end{pmatrix},
    \label{eq:rewritemotion}
\end{align}
by defining $\beta = 2PL/[c(1 - g_1g_2)]$, $g_i = 1 - L/R_i$.
$g_i$ is determined by the geometry of the cavity and generally called the $g$ factor.

Hereafter, we consider a case where the mirror 2 is much heavier than the mirror 1, and the mechanical restoring torque of the mirror 1 is much smaller than that of the mirror 2.
This assumption is practical for actual experiments.
This optomechanical system can be a sensitive force sensor by using the lighter mirror as a test mass.
At the same time, we can control the cavity length to maintain the resonance by attaching actuators on the larger (heavier) mirror; $I_1 \ll I_2$, and $K_1 \ll K_2$.
In this case, the diagonalization of Eq.~(\ref{eq:rewritemotion}) indicates the resonant frequency of the differential mode is
\begin{align}
    \omega_\mathrm{diff} \simeq \sqrt{\frac{K_1 - \beta g_2}{I_1}}.
    \label{eq:resfreqdiff}
\end{align}
The diagonalization of Eq.~(\ref{eq:rewritemotion}) derives two decoupled modes.
Here, the eigen mode where the two mirrors rotate in the same direction is named the differential mode.
On the other hand, the eigen mode in which the mirrors rotate in the opposite directions is named the common mode.

When the lighter mirror is flat ($g_1 = 1$), $g_2$ should be positive to satisfy the optical cavity condition of $0 < g_1g_2 <1$.
In this case, the resonant frequency rapidly goes to zero with the increase of the intracavity laser power, which causes an angular instability.
On the other hand, we can avoid the instability and even can stiffen the differential mode in the negative-$g$ regime.
Equation~(\ref{eq:rewritemotion}) also indicates the resonant frequency of the common mode decreases as
\begin{align}
    \omega_\mathrm{com} \simeq \sqrt{\frac{K_2 + \beta(1 - g^2)/g}{I_2}}.
\end{align}
Here, we assume the two curvatures of the mirrors are identical ($R_1 = R_2 = R$, $g_1 = g_2 = g$) for simplicity.
The mechanical resonant frequency of the common mode can be high enough by using a heavy enough mirror for the mirror 2.
Therefore, the decrease of the resonant frequency due to the radiation pressure torque can be ignored.
In other words, the radiation pressure torque will not make the common mode unstable when one mirror is much heavier than the other.
The tolerable intracavity power can be increased by increasing only the mass of one mirror.
Thus, this configuration allows trapping of a small mirror.

\begin{figure}
    \centering
    \includegraphics[width=\columnwidth,clip]{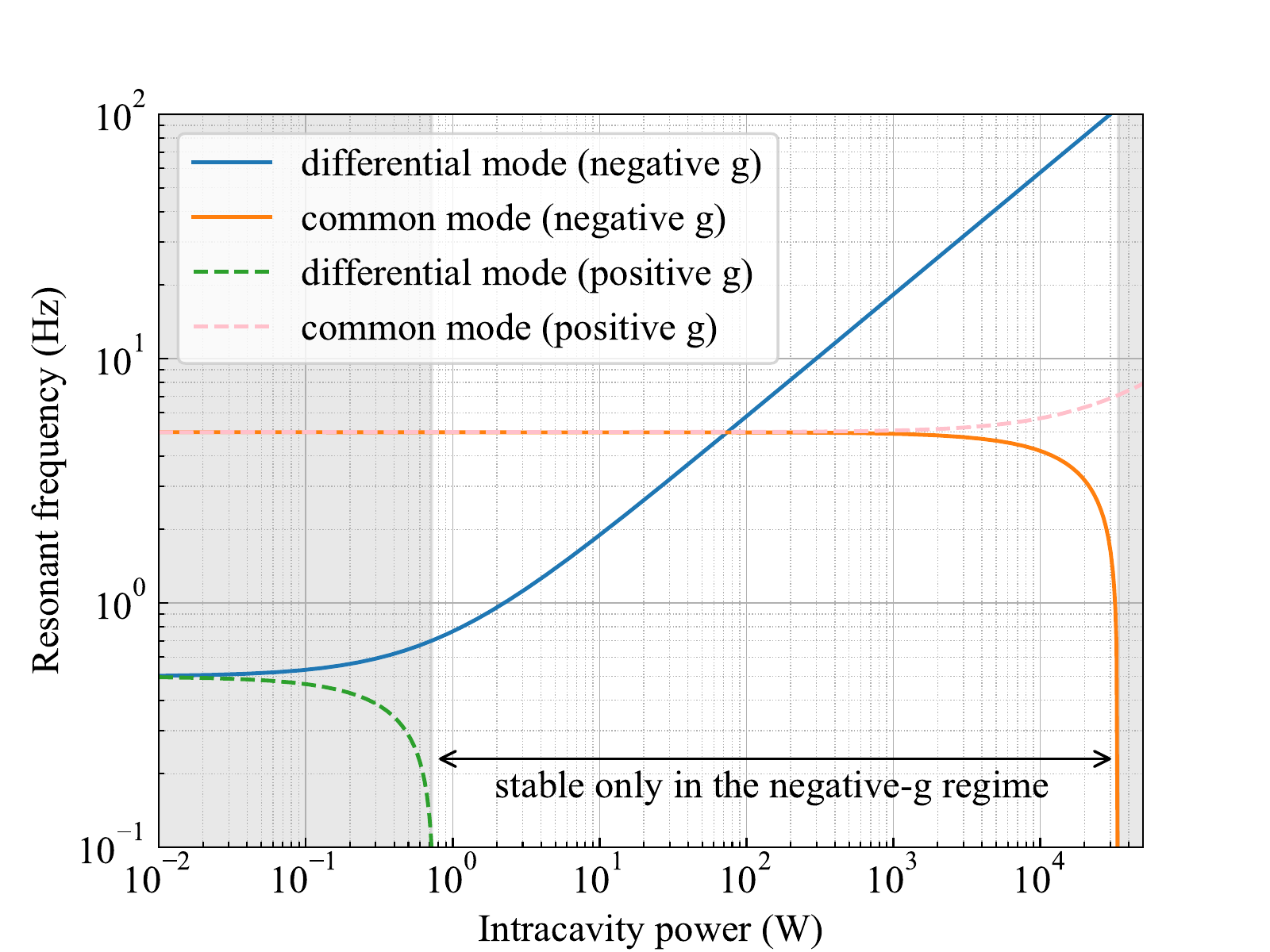}
    \caption{Dependence of the resonant frequencies of differential and common modes.
    The negative resonant frequency implies that the mode is unstable.
    For comparison, both the negative-$g$ regime ($g_1 = g_2 = -0.1$) and the positive-$g$ regime ($g_1 = g_2 = +0.1$) cases are plotted; other parameters are described in the main text.
    In the range between 0.72~W and 34~kW of the intracavity power, only the negative-$g$ cavity is stable.}
    \label{fig:theoretical}
\end{figure}

Figure~\ref{fig:theoretical}. shows the dependence of the resonant frequency on the intracavity power.
As for this plot, we use similar parameters of our experimental setup as follows: mirror masses of $m_1 = 10~\mathrm{mg}$ and $m_2 = 10~\mathrm{g}$, mirror radii of $r_1 = 1.5~\mathrm{mm}$ and $r_2 = 10~\mathrm{mm}$, (momentum of intertia of $I_1 = 5.6\times 10^{-12}~\mathrm{kg~m^2}$ and $I_2 = 2.5\times 10^{-7\textbf{}}~\mathrm{kg~m^2}$,) mechanical resonant frequency of $\omega_1/(2\pi) = 0.5~\mathrm{Hz}$ and $\omega_2/(2\pi) = 5~\mathrm{Hz}$, and the cavity length of $L = 1.1R = 11~\mathrm{cm}~(g = -0.1)$.
For comparison, we also plot the resonant frequencies of the differential and common modes for the case that the $g$ factors of the cavity mirrors are positive ($g_1 =g_2 = +0.1$).
The cavity in the negative-$g$ regime is tolerant over the intracavity power of 10~kW, while the cavity in the positive-$g$ regime is unstable just over 0.72~W.
\begin{figure*}
    \centering
    \includegraphics[width=2\columnwidth,clip]{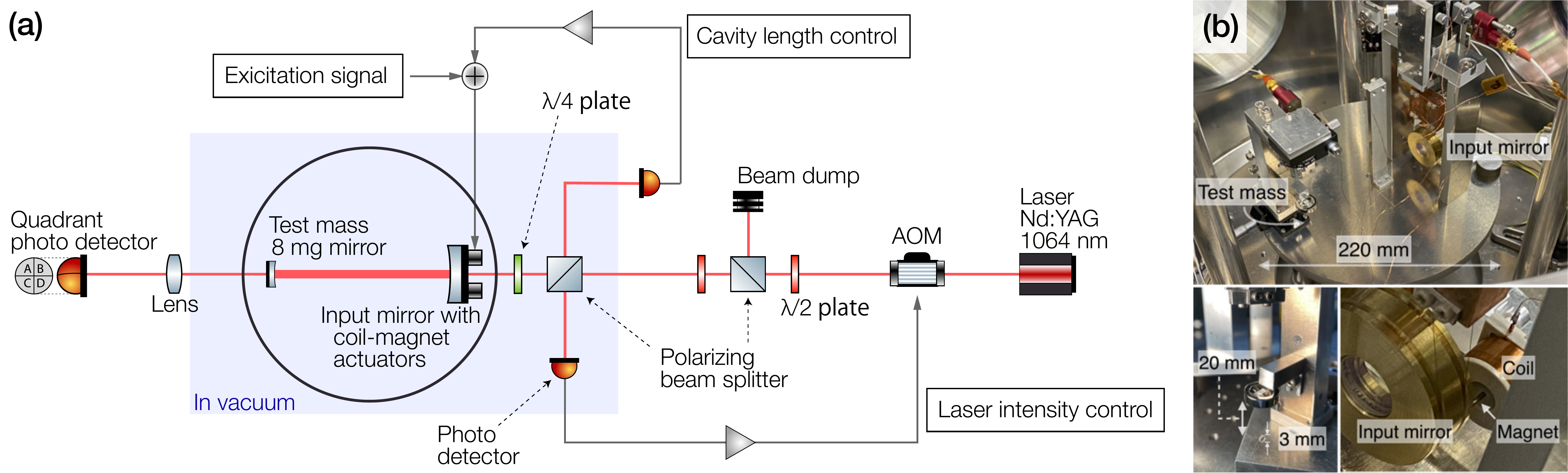}
    \caption{(a)Schematic drawing of the experimental setup.
                The test mass is an 8~mg mirror, which is suspended as a double pendulum.
                The test mass and the input mirror consist of the main cavity on the platform in vacuum chamber.
                The laser source is an Nd:YAG laser, and the wavelength is 1064~nm.
                The cavity length is feedback controlled for the continuous resonance.
                The laser intensity fluctuation is also suppressed by the feedback control with an acousto-optic modulator.
                The beam spot on the test mass is monitored with a quadrant photo detector for the transfer function measurement.
             (b)Photographs of the main cavity on the platform.
                The platform is suspended with springs as a double pendulum to isolate the main cavity from seismic vibration.
                The test mass is focused on the lower left side. The input mirror and the coil magnet actuator on it is focused on the lower right side.}
    \label{fig:setup}
\end{figure*}

\section{\label{sec:method}Experimental demonstration}
We experimentally demonstrate that our trapping configuration works properly.
% for a suspended mirror in the rotational degree of freedom with our geometrical configuration and to demonstrate the new trapping scheme is valid.
As mentioned in the previous section, our configuration utilizes the radiation pressure of the laser light inside a linear cavity.
Thus, the restoring torque due to the radiation pressure also increases as the intracavity power increases.
We design our experiment to observe this increase of the restoring torque.

The experimental setup is shown in Fig.~\ref{fig:setup}.
The instability due to the radiation pressure of the laser light inside the cavity will be an issue when the radiation pressure torque is dominant.
To realize the predominance of the radiation pressure, we build a linear cavity with a tiny mirror of 8~mg (0.5~mm thick with a diameter 3~mm) as the test mass.
This tiny mirror is suspended with a thin carbon fiber (6~$\mu$m in diameter and 2~cm long) to lower the mechanical restoring torque.
The $Q$ value of a single pendulum with this carbon fiber is measured to be $Q \sim 8\times 10^4$ at the mechanical resonant frequency, 3~Hz.
The test mass is suspended as a double pendulum via the intermediate mass for isolation from the seismic motion.
The input mirror is much heavier.
Its mass is 60~g.
Two coil-magnet actuators are attached to the input mirror to apply a force and a torque.
The radii of the curvature of the mirrors are 10~cm, and the cavity length is $11.0~\pm 0.3$~cm.
Thus, the cavity is in the negative-$g$ regime of $g = -0.1$.
The finesse of the cavity is measured to be $(3.0 \pm 0.3)\times 10^3$.
The cavity is built on a platform board; the platform is also suspended as a double pendulum in a vacuum chamber.
The pressure is kept about 1~Pa to suppress acoustic disturbances.

The cavity length is feedback controlled to resonate continuously.
We use a side locking to keep the cavity at a detuned point near half of the resonant peak.
The displacement of the mirror from the control point is sensed by the reflected-light power that is monitored by the photo detector.
The error signal is obtained by comparing the output signal from the photo detector with a constant voltage.
The error signal is filtered and sent to the coil-magnet actuator attached to the input mirror.
The transmitted light from the cavity is monitored by a quadrant photo detector, as described in Fig.~\ref{fig:setup}.
The half of the injected laser beam is also monitored for the laser intensity control.
By sending the feedback signal to the acousto-optic modulator (AOM), we suppress the laser intensity fluctuation.

To show the radiation pressure works as a positive restoring torque, we evaluate the resonant frequency that is described in Eq.~(\ref{eq:resfreqdiff}).
The resonant frequency is determined by the transfer function of the rotational motion of the mirrors.
We apply torque to the input mirror by injecting differential sine-wave signals into the coil-magnet actuators on the input mirror~\ref{fig:setup}.
Then, the test mass is also swung via the radiation pressure inside the cavity.
The rotation of the test mass results in the changes of the beam spot on the test mass because the cavity axis changes.
We observe the transmitted light from the cavity by a quadrant photo detector to detect the change in the beam spot.
A convex lens is placed halfway between the test mass and the quadrant photo detector to measure only the change in the beam spot without being affected by the change in the cavity axis.
The distance between the lens and the test mass (the quadrant photo detector) is twice as long as the focal length of the lens.
The transfer function from the injected signal to quadrant photo detector signal includes the transfer function of the suspended mirrors~\cite{Nagano:2016wcg}.
Thus, the transfer function has the characteristic form of the resonant peak.
We fit the transfer function to estimate the resonant frequency.
Note that the cavity length is feedback controlled to keep the resonance of the cavity during this transfer function measurement.

\section{\label{sec:result}Result}
\begin{figure}
    \centering
    \includegraphics[width=\columnwidth,clip]{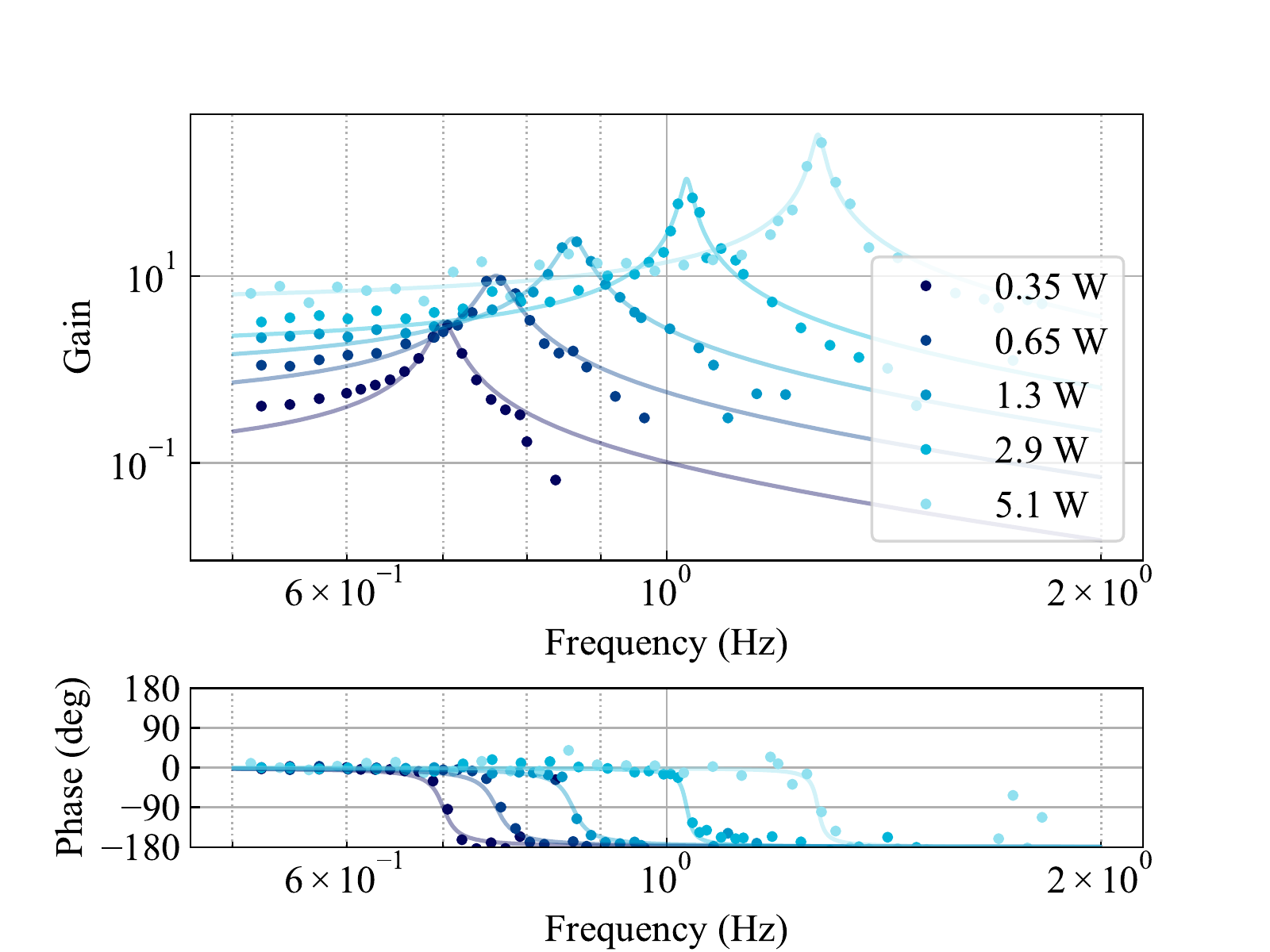}
    \caption{The measured transfer functions of the five measurements of different intracavity powers. The points are the measured data, and the lines represent the fittings. The peaks and the phase flips indicate the resonant points. By the fitting, we estimate the resonant frequencies.}
    \label{fig:transfunc}
\end{figure}

The measured transfer functions from the excitation to the quadrant photo detector signal are plotted in Fig.~\ref{fig:transfunc}.
We measure them with five different intracavity powers.
The peaks and the phase flips due to the resonance of the test mass are observed in each measurement.
The intracavity power in each measurement is estimated by the transmitted light power.
We fit the gain of the transfer functions to determine the resonant frequency.
The fitted parameters are resonant frequency, damping ratio of the resonance, and the overall gain factor.
The fitted curves are also plotted in Fig.~\ref{fig:transfunc}.
The uncertainty in the intracavity power is dominated by the fluctuation in the power of the transmitted light.
The fluctuation in the transmitted light power is at the frequency of the excitation signal.
Thus, the fluctuation would be due to the misalignment when the mirror is swung.

\begin{figure}
    \centering
    \includegraphics[width=\columnwidth,clip]{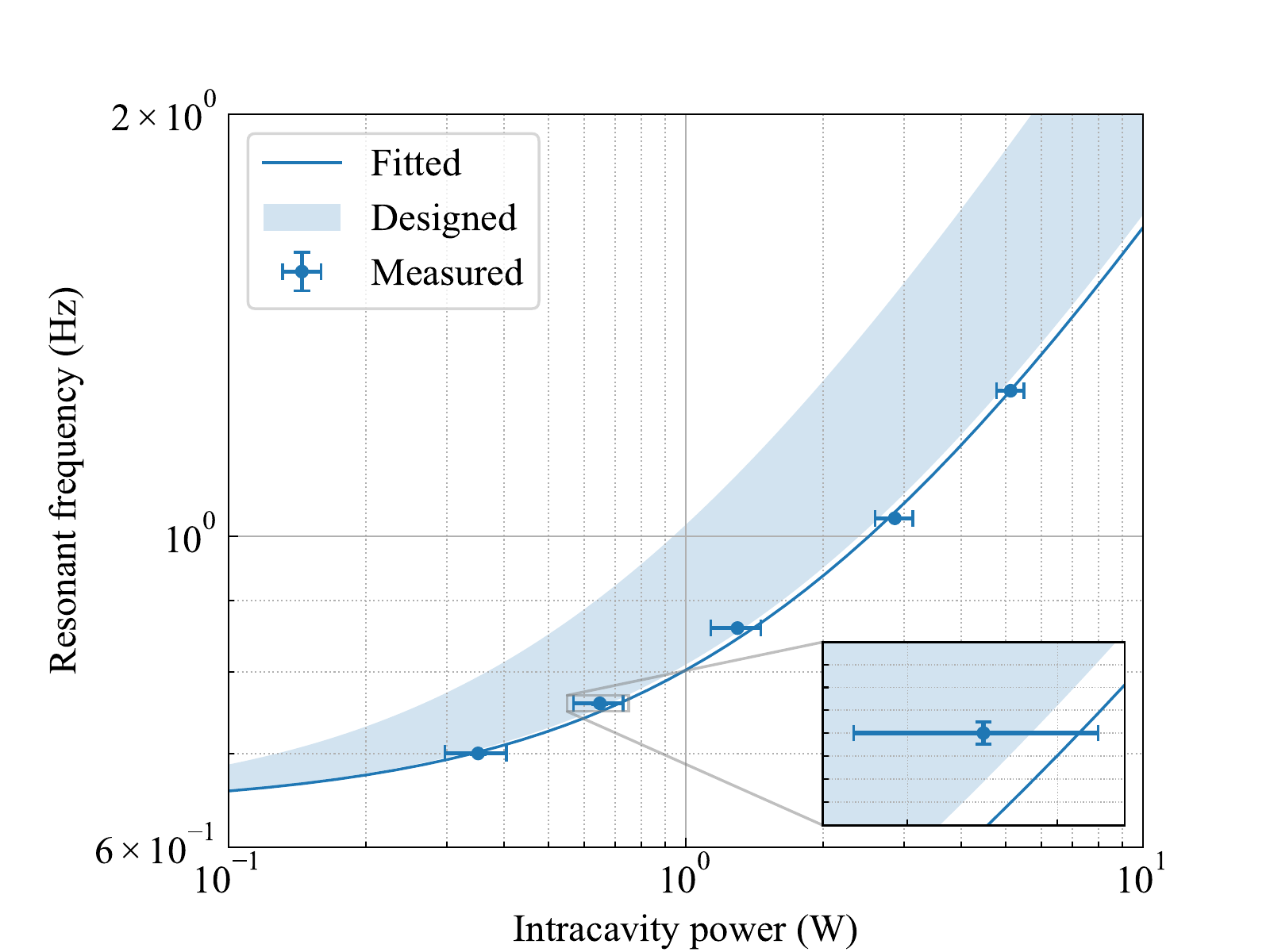}
    \caption{The resonant frequencies of the differential mode in the rotational degree of freedom. The shaded region represents the theoretically predicted values with the width corresponding to the uncertainties in mirror reflectivities and the cavity length.}
    \label{fig:resfreq}
\end{figure}
We show the dependence of the resonant frequency on the intracavity power in Fig.~\ref{fig:resfreq}.
The uncertainty of the resonant frequency is estimated from the fit of the transfer function measurement.
We also show the predicted region from the theoretical calculation using Eq.~(\ref{eq:resfreqdiff}) with the parameters of the optics.
The width of the region corresponds to the uncertainty of the design reflectivities of the mirrors and the uncertainty of the cavity length.
The measured dependency is consistent with the theoretical prediction.

% \subsection{\label{subsec:os_coupling}Effect of the optical spring}
We note that the longitudinal optical spring effect does not affect our results since the beam spot on the test mass mirror is precisely adjusted to the center.
When the beam spot is off the center of the mirror, the longitudinal optical spring also acts as a restoring torque.
However, the deviation of the beam spot from the center of the test mass must be smaller than 2~$\mu$m with our experimental parameters.
Otherwise, the constant radiation pressure rotates the test mass mirror and breaks the cavity locking.
When the deviation is smaller than 2~$\mu$m, the restoring torque from the longitudinal optical spring is smaller than several $10^{-11}$~Nm/rad.
This is smaller than the original restoring torque of the pendulum's suspension wire.
Furthermore, the trapping potential by the longitudinal optical spring is suppressed by the feedback controlling of the cavity length, and thus, its effect is further minimized.
Therefore, we conclude that the effect of the coupling of the longitudinal optical spring is negligible.

\section{\label{sec:discussion}Discussion}
One of the essential applications of our trapping scheme is the observation of the quantum radiation pressure fluctuation for testing macroscopic quantum mechanics; the quantum radiation pressure fluctuation is a radiation pressure fluctuation due to the quantum fluctuation of the photon number of light.
As mentioned in the Introduction, one attracting target of milligram scale optomechanical systems is testing macroscopic quantum mechanics.
A proposed experimental test of macroscopic quantum mechanics requires to prepare a conditional state~\cite{Chen:2013sgs}.
To set an optomechanical system in a conditional state, the quantum radiation pressure fluctuation needs to dominate force noises~\cite{Michimura:2020yvn}.
For the observation of quantum radiation pressure fluctuation, the combination of a light mirror and a high power laser beam is preferable.
The high power laser beam enhances the quantum radiation pressure fluctuation itself, and a light mirror enhances the displacement of the mirror that is sensed by the interference.
Therefore, our trapping system is suitable because it overcomes the Sidles-Sigg instability, which limited the optimal sensitivities in the previous milligram and gram scale experiments~\cite{Corbitt:2007spn,Neben:2012edt,Kelley:2015axa,Sakata:2010zz,Matsumoto:2013sua,Matsumoto:2014fda,Komori:2019zlg}.%; at the scale, an active feedback control is not always available since an actuator cannot be attached on the oscillator.
% \subsection{\label{subsec:applicationQRPF}Applicability for observation of the quantum radiation pressure fluctuation}

We estimate the sensitivity of our experimental setup to discuss the feasibility of observing the quantum radiation pressure fluctuation.
The calculation reveals that the quantum radiation pressure fluctuation will be dominant with the intracavity power of over 14~W (see the Appendix~\ref{sec:appendix} for more details).
Since our trapping scheme overcomes the limitation of the Sidles-Sigg instability, the cavity can accumulate 30~kW or more power inside, according to our measurement result.
Therefore, we conclude that our system can realize the observation of the quantum radiation pressure fluctuation, though technical classical noises should be well suppressed.

\section{\label{sec:conclusion}Conclusion}
We propose a configuration to trap the rotational motions of the suspended mirrors in a linear cavity.
By operating a linear cavity in the negative-$g$ regime and using unbalanced-mass mirrors, the two rotational modes of the cavity mirrors are stable with the radiation pressure inside the cavity.
Furthermore, we demonstrate an experimental validation of the trapping by building a linear cavity with an 8~mg mirror.
We observe the rotational restoring torque on the mirror increases as the intracavity power increases.
The behavior is consistent with theoretical prediction.
Therefore, we confirm that the 8~mg mirror obtains the positive restoring torque originating from the radiation pressure of the inside laser beam.

We also discuss the possibility of observing quantum radiation pressure fluctuation by using our trapping scheme.
The calculation shows that the quantum radiation pressure fluctuation can be observed with the realistic parameters that are identical or similar to our realized experiment.
Considering the successful result of the trapping together, we confirm that the cavity can accumulate enough large laser power to enhance the quantum radiation pressure fluctuation to observe it.
Thus, this work is a crucial step towards testing macroscopic quantum mechanics, while the configuration is also applicable to a broad range of optomechanical systems.

\begin{acknowledgments}
We thank Yutaro Enomoto and Ooi Ching Pin for fruitful discussions.
% We also thank OptoSigma company for producing the mg scale curved mirror.
This work was supported by a Grant-in-Aid for Challenging Research Exploratory Grant No. 18K18763 from the Japan Society for the Promotion of Science (JSPS), by JST CREST Grant No. JPMJCR1873, and by MEXT Quantum LEAP Flagship Program (MEXT Q-LEAP) Grant No. JPMXS0118070351.
T.K. is supported by KAKENHI Grant No. 19J21861 from the JSPS.
\end{acknowledgments}

\appendix
\section{\label{sec:appendix}Calculation of the sensitivity for our linear cavity}
\begin{figure}
    \centering
    \includegraphics[width=\columnwidth,clip]{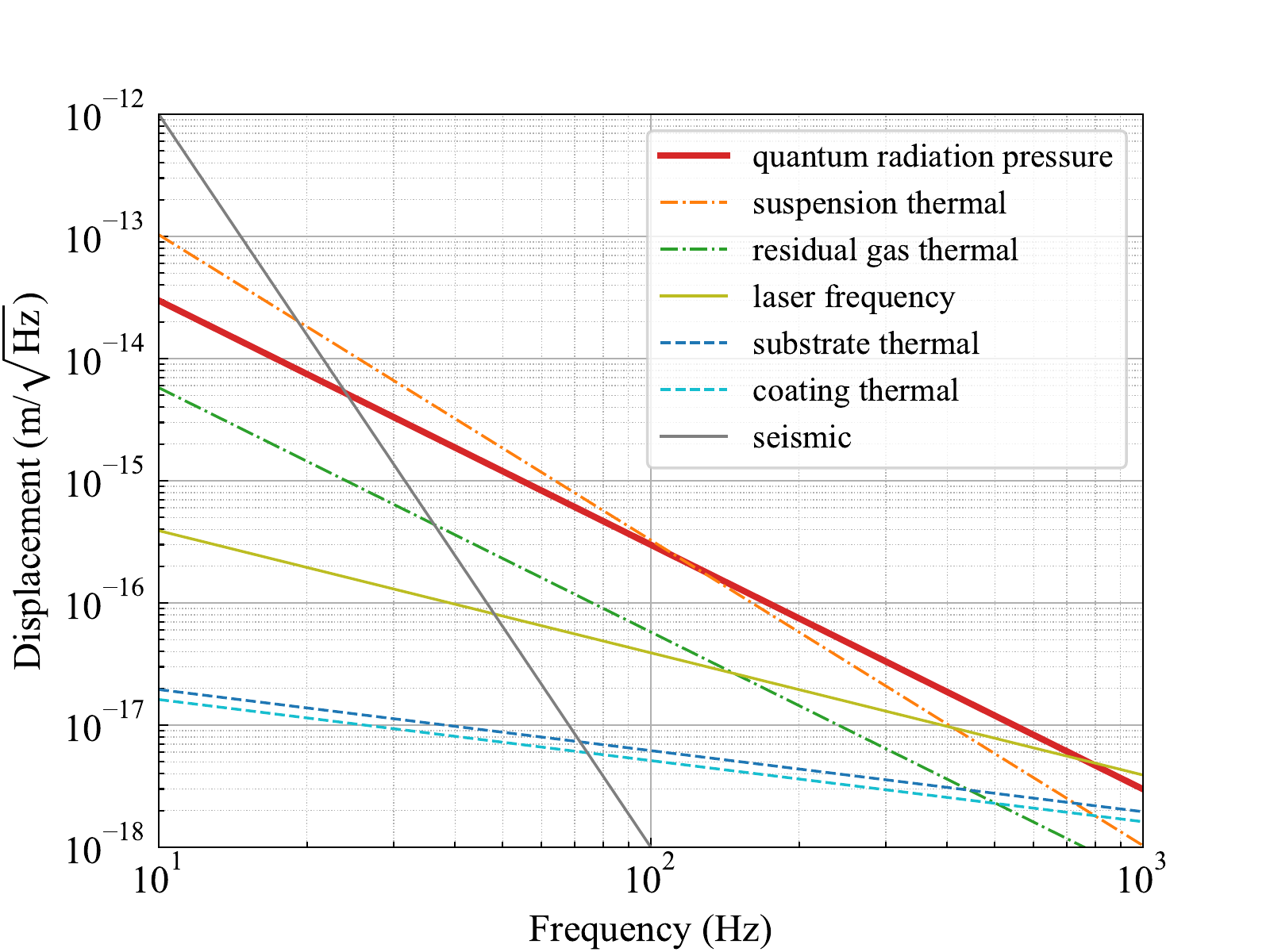}
    \caption{Design sensitivity for observing quantum radiation pressure fluctuation. To calculate the sensitivity, we assume the identical or similar parameters of our experiment. The signal to noise ratio of quantum radiation pressure fluctuation to the sum of the other classical noises is over 1 in the range between 130~Hz and 600~Hz.}
    \label{fig:designsensitivity}
\end{figure}

We estimate the sensitivity of our linear cavity to discuss the feasibility of the observation of the quantum radiation pressure fluctuation.
We calculate the sensitivity with parameters shown in Table~\ref{tab:parameters}.
\begin{table}
    \caption{Parameters for the calculation of the design sensitivity towards observation of the quantum radiation pressure fluctuation. The parameters are based on our realized experimental setup.}
    \centering
    \begin{ruledtabular}
    \begin{tabular}{l l r}
        Test mass & & \\
        Mass & & 8 mg \\
        Diameter & & 3 mm\\
        Radius of curvature & & 100 mm \\
        Reflectivity & & 99.99\% \\
        $Q$ value & & $10^5$ \\
        Beam radius & & 0.21~mm \\
        Young's modulus & Substrate & 73 GPa \\
        & SiO$_2$ & 73 GPa \\        
        & TiO$_2$:Ta$_2$O$_5$ & 140 GPa \\
        Poisson ratio & Substrate & 0.17 \\
        & SiO$_2$ & 0.17 \\
        & TiO$_2$:Ta$_2$O$_5$ & 0.28 \\
        Loss angle & Substrate & $1\times 10^{-5}$ \\
        & SiO$_2$ & $1\times 10^{-4}$ \\
        & TiO$_2$:Ta$_2$O$_5$ & $4\times 10^{-4}$ \\
        Refractive index & Substrate & 1.45 \\
        & SiO$_2$ & 1.45 \\
        & TiO$_2$:Ta$_2$O$_5$ & 2.07 \\
        
        \colrule
        Input mirror & & \\
        Mass & & 60 g \\
        Radius of curvature & & 100 mm \\
        Reflectivity & & 99.9\% \\

        \colrule
        Cavity & & \\
        Cavity length & & 110 mm \\
        Finesse & & 5000 \\
        Intracavity power & & 14~W\\

        \colrule
        Laser & & \\
        Wavelength & & 1064 nm \\
        Input power & & 10 mW \\
        Frequency noise & & $10~\text{Hz}/f$ Hz/$\sqrt{\mathrm{Hz}}$ \\
        \colrule
        Temperature & & 300 K \\
        Air pressure & & $10^{-4}$ Pa
    \end{tabular}
    \end{ruledtabular}
    \label{tab:parameters}
\end{table}
As shown in Fig.~\ref{fig:designsensitivity}, the design sensitivity reaches quantum radiation pressure fluctuation with the input laser power of 10~mW.
To estimate the mirror thermal noises, we assume that the finesse of the cavity is 5000, which is realized with a coating reflectivity of 99.99\% on the 8~mg mirror and a coating reflectivity of 99.9\% on the input mirror.
These reflectivities are the specification values of the mirrors of our experiment.
The mirror substrate is made of fused silica, and the coatings of the mirrors are multilayers of SiO$_2$ and TiO$_2$:Ta$_2$O$_5$.
The substrate and coating thermal noises are calculated based on the calculation in~\cite{Levin:1997kv, Bondu:1998sm, Villar:2010kf}.
It is known that the spectrum of seismic noise is typically $10^{-7}/f^2~\mathrm{m/\sqrt{Hz}}$ above 0.1~Hz~\cite{Shoemaker:1987ft}.
Because the platform of our experiment is suspended as a double pendulum with springs, we assume suppression of the seismic noise by the factor of $1/f^4$ to calculate the design sensitivity.
The air pressure should be below $10^{-4}$~Pa so that the thermal noise of the residual gas~\cite{Saulson:1990jc} is small enough.

For the sensitivity estimation, we imply the active stabilization of the laser.
The laser frequency noise of an Nd:YAG laser is typically around 10-100~$\mathrm{Hz/\sqrt{Hz}}$ in the region of interest~\cite{Bondu:1996rnn,Conti:2003}.
By a frequency locking to a stable external cavity, the required frequency noise level is achievable~\cite{Bondu:1996rnn,Conti:2003}.
The laser intensity noise can be reduced with a feedback control to the level of quantum noise for the required laser power of 10~mW~\cite{Conti:2000nhd}.

% \nocite{*}

\bibliography{mqm}% Produces the bibliography via BibTeX.

\begin{thebibliography}{40}%
\makeatletter
\providecommand \@ifxundefined [1]{%
 \@ifx{#1\undefined}
}%
\providecommand \@ifnum [1]{%
 \ifnum #1\expandafter \@firstoftwo
 \else \expandafter \@secondoftwo
 \fi
}%
\providecommand \@ifx [1]{%
 \ifx #1\expandafter \@firstoftwo
 \else \expandafter \@secondoftwo
 \fi
}%
\providecommand \natexlab [1]{#1}%
\providecommand \enquote  [1]{``#1''}%
\providecommand \bibnamefont  [1]{#1}%
\providecommand \bibfnamefont [1]{#1}%
\providecommand \citenamefont [1]{#1}%
\providecommand \href@noop [0]{\@secondoftwo}%
\providecommand \href [0]{\begingroup \@sanitize@url \@href}%
\providecommand \@href[1]{\@@startlink{#1}\@@href}%
\providecommand \@@href[1]{\endgroup#1\@@endlink}%
\providecommand \@sanitize@url [0]{\catcode `\\12\catcode `\$12\catcode
  `\&12\catcode `\#12\catcode `\^12\catcode `\_12\catcode `\%12\relax}%
\providecommand \@@startlink[1]{}%
\providecommand \@@endlink[0]{}%
\providecommand \url  [0]{\begingroup\@sanitize@url \@url }%
\providecommand \@url [1]{\endgroup\@href {#1}{\urlprefix }}%
\providecommand \urlprefix  [0]{URL }%
\providecommand \Eprint [0]{\href }%
\providecommand \doibase [0]{https://doi.org/}%
\providecommand \selectlanguage [0]{\@gobble}%
\providecommand \bibinfo  [0]{\@secondoftwo}%
\providecommand \bibfield  [0]{\@secondoftwo}%
\providecommand \translation [1]{[#1]}%
\providecommand \BibitemOpen [0]{}%
\providecommand \bibitemStop [0]{}%
\providecommand \bibitemNoStop [0]{.\EOS\space}%
\providecommand \EOS [0]{\spacefactor3000\relax}%
\providecommand \BibitemShut  [1]{\csname bibitem#1\endcsname}%
\let\auto@bib@innerbib\@empty
%</preamble>
\bibitem [{\citenamefont {Aspelmeyer}\ \emph {et~al.}(2014)\citenamefont
  {Aspelmeyer}, \citenamefont {Kippenberg},\ and\ \citenamefont
  {Marquardt}}]{Aspelmeyer:2013lha}%
  \BibitemOpen
  \bibfield  {author} {\bibinfo {author} {\bibfnamefont {M.}~\bibnamefont
  {Aspelmeyer}}, \bibinfo {author} {\bibfnamefont {T.~J.}\ \bibnamefont
  {Kippenberg}},\ and\ \bibinfo {author} {\bibfnamefont {F.}~\bibnamefont
  {Marquardt}},\ }\bibfield  {title} {\bibinfo {title} {{Cavity
  Optomechanics}},\ }\href {https://doi.org/10.1103/RevModPhys.86.1391}
  {\bibfield  {journal} {\bibinfo  {journal} {Rev. Mod. Phys.}\ }\textbf
  {\bibinfo {volume} {86}},\ \bibinfo {pages} {1391} (\bibinfo {year}
  {2014})},\ \Eprint {https://arxiv.org/abs/1303.0733} {arXiv:1303.0733
  [cond-mat.mes-hall]} \BibitemShut {NoStop}%
\bibitem [{\citenamefont {Chen}(2013)}]{Chen:2013sgs}%
  \BibitemOpen
  \bibfield  {author} {\bibinfo {author} {\bibfnamefont {Y.}~\bibnamefont
  {Chen}},\ }\bibfield  {title} {\bibinfo {title} {{Macroscopic Quantum
  Mechanics: Theory and Experimental Concepts of Optomechanics}},\ }\href
  {https://doi.org/10.1088/0953-4075/46/10/104001} {\bibfield  {journal}
  {\bibinfo  {journal} {J. Phys. B}\ }\textbf {\bibinfo {volume} {46}},\
  \bibinfo {pages} {104001} (\bibinfo {year} {2013})},\ \Eprint
  {https://arxiv.org/abs/1302.1924} {arXiv:1302.1924 [quant-ph]} \BibitemShut
  {NoStop}%
\bibitem [{\citenamefont {Miao}\ \emph {et~al.}(2020)\citenamefont {Miao},
  \citenamefont {Martynov}, \citenamefont {Yang},\ and\ \citenamefont
  {Datta}}]{Miao:2019pxw}%
  \BibitemOpen
  \bibfield  {author} {\bibinfo {author} {\bibfnamefont {H.}~\bibnamefont
  {Miao}}, \bibinfo {author} {\bibfnamefont {D.}~\bibnamefont {Martynov}},
  \bibinfo {author} {\bibfnamefont {H.}~\bibnamefont {Yang}},\ and\ \bibinfo
  {author} {\bibfnamefont {A.}~\bibnamefont {Datta}},\ }\bibfield  {title}
  {\bibinfo {title} {{Quantum correlation of light mediated by gravity}},\
  }\href {https://doi.org/10.1103/PhysRevA.101.063804} {\bibfield  {journal}
  {\bibinfo  {journal} {Phys. Rev. A}\ }\textbf {\bibinfo {volume} {101}},\
  \bibinfo {pages} {063804} (\bibinfo {year} {2020})},\ \Eprint
  {https://arxiv.org/abs/1901.05827} {arXiv:1901.05827 [quant-ph]} \BibitemShut
  {NoStop}%
\bibitem [{\citenamefont {Kafri}\ \emph {et~al.}(2014)\citenamefont {Kafri},
  \citenamefont {Taylor},\ and\ \citenamefont {Milburn}}]{Kafri:2014zsa}%
  \BibitemOpen
  \bibfield  {author} {\bibinfo {author} {\bibfnamefont {D.}~\bibnamefont
  {Kafri}}, \bibinfo {author} {\bibfnamefont {J.~M.}\ \bibnamefont {Taylor}},\
  and\ \bibinfo {author} {\bibfnamefont {G.~J.}\ \bibnamefont {Milburn}},\
  }\bibfield  {title} {\bibinfo {title} {{A classical channel model for
  gravitational decoherence}},\ }\href
  {https://doi.org/10.1088/1367-2630/16/6/065020} {\bibfield  {journal}
  {\bibinfo  {journal} {New J. Phys.}\ }\textbf {\bibinfo {volume} {16}},\
  \bibinfo {pages} {065020} (\bibinfo {year} {2014})},\ \Eprint
  {https://arxiv.org/abs/1401.0946} {arXiv:1401.0946 [quant-ph]} \BibitemShut
  {NoStop}%
\bibitem [{\citenamefont {Schm\"ole}\ \emph {et~al.}(2016)\citenamefont
  {Schm\"ole}, \citenamefont {Dragosits}, \citenamefont {Hepach},\ and\
  \citenamefont {Aspelmeyer}}]{Schmole:2016mde}%
  \BibitemOpen
  \bibfield  {author} {\bibinfo {author} {\bibfnamefont {J.}~\bibnamefont
  {Schm\"ole}}, \bibinfo {author} {\bibfnamefont {M.}~\bibnamefont
  {Dragosits}}, \bibinfo {author} {\bibfnamefont {H.}~\bibnamefont {Hepach}},\
  and\ \bibinfo {author} {\bibfnamefont {M.}~\bibnamefont {Aspelmeyer}},\
  }\bibfield  {title} {\bibinfo {title} {{A micromechanical proof-of-principle
  experiment for measuring the gravitational force of milligram masses}},\
  }\href {https://doi.org/10.1088/0264-9381/33/12/125031} {\bibfield  {journal}
  {\bibinfo  {journal} {Class. Quant. Grav.}\ }\textbf {\bibinfo {volume}
  {33}},\ \bibinfo {pages} {125031} (\bibinfo {year} {2016})},\ \Eprint
  {https://arxiv.org/abs/1602.07539} {arXiv:1602.07539 [physics.ins-det]}
  \BibitemShut {NoStop}%
\bibitem [{\citenamefont {Abbott}\ \emph {et~al.}(2016)\citenamefont {Abbott}
  \emph {et~al.}}]{LIGOScientific:2016aoc}%
  \BibitemOpen
  \bibfield  {author} {\bibinfo {author} {\bibfnamefont {B.~P.}\ \bibnamefont
  {Abbott}} \emph {et~al.} (\bibinfo {collaboration} {LIGO Scientific,
  Virgo}),\ }\bibfield  {title} {\bibinfo {title} {{Observation of
  Gravitational Waves from a Binary Black Hole Merger}},\ }\href
  {https://doi.org/10.1103/PhysRevLett.116.061102} {\bibfield  {journal}
  {\bibinfo  {journal} {Phys. Rev. Lett.}\ }\textbf {\bibinfo {volume} {116}},\
  \bibinfo {pages} {061102} (\bibinfo {year} {2016})},\ \Eprint
  {https://arxiv.org/abs/1602.03837} {arXiv:1602.03837 [gr-qc]} \BibitemShut
  {NoStop}%
\bibitem [{\citenamefont {Marletto}\ and\ \citenamefont
  {Vedral}(2017)}]{Marletto:2017kzi}%
  \BibitemOpen
  \bibfield  {author} {\bibinfo {author} {\bibfnamefont {C.}~\bibnamefont
  {Marletto}}\ and\ \bibinfo {author} {\bibfnamefont {V.}~\bibnamefont
  {Vedral}},\ }\bibfield  {title} {\bibinfo {title} {{Gravitationally-induced
  entanglement between two massive particles is sufficient evidence of quantum
  effects in gravity}},\ }\href
  {https://doi.org/10.1103/PhysRevLett.119.240402} {\bibfield  {journal}
  {\bibinfo  {journal} {Phys. Rev. Lett.}\ }\textbf {\bibinfo {volume} {119}},\
  \bibinfo {pages} {240402} (\bibinfo {year} {2017})},\ \Eprint
  {https://arxiv.org/abs/1707.06036} {arXiv:1707.06036 [quant-ph]} \BibitemShut
  {NoStop}%
\bibitem [{\citenamefont {Bose}\ \emph {et~al.}(2017)\citenamefont {Bose},
  \citenamefont {Mazumdar}, \citenamefont {Morley}, \citenamefont {Ulbricht},
  \citenamefont {Toro\ifmmode~\check{s}\else \v{s}\fi{}}, \citenamefont
  {Paternostro}, \citenamefont {Geraci}, \citenamefont {Barker}, \citenamefont
  {Kim},\ and\ \citenamefont {Milburn}}]{Bose:2017nin}%
  \BibitemOpen
  \bibfield  {author} {\bibinfo {author} {\bibfnamefont {S.}~\bibnamefont
  {Bose}}, \bibinfo {author} {\bibfnamefont {A.}~\bibnamefont {Mazumdar}},
  \bibinfo {author} {\bibfnamefont {G.~W.}\ \bibnamefont {Morley}}, \bibinfo
  {author} {\bibfnamefont {H.}~\bibnamefont {Ulbricht}}, \bibinfo {author}
  {\bibfnamefont {M.}~\bibnamefont {Toro\ifmmode~\check{s}\else \v{s}\fi{}}},
  \bibinfo {author} {\bibfnamefont {M.}~\bibnamefont {Paternostro}}, \bibinfo
  {author} {\bibfnamefont {A.~A.}\ \bibnamefont {Geraci}}, \bibinfo {author}
  {\bibfnamefont {P.~F.}\ \bibnamefont {Barker}}, \bibinfo {author}
  {\bibfnamefont {M.~S.}\ \bibnamefont {Kim}},\ and\ \bibinfo {author}
  {\bibfnamefont {G.}~\bibnamefont {Milburn}},\ }\bibfield  {title} {\bibinfo
  {title} {{Spin Entanglement Witness for Quantum Gravity}},\ }\href
  {https://doi.org/10.1103/PhysRevLett.119.240401} {\bibfield  {journal}
  {\bibinfo  {journal} {Phys. Rev. Lett.}\ }\textbf {\bibinfo {volume} {119}},\
  \bibinfo {pages} {240401} (\bibinfo {year} {2017})},\ \Eprint
  {https://arxiv.org/abs/1707.06050} {arXiv:1707.06050 [quant-ph]} \BibitemShut
  {NoStop}%
\bibitem [{\citenamefont {Belenchia}\ \emph {et~al.}(2018)\citenamefont
  {Belenchia}, \citenamefont {Wald}, \citenamefont {Giacomini}, \citenamefont
  {Castro-Ruiz}, \citenamefont {Brukner},\ and\ \citenamefont
  {Aspelmeyer}}]{Belenchia:2018szb}%
  \BibitemOpen
  \bibfield  {author} {\bibinfo {author} {\bibfnamefont {A.}~\bibnamefont
  {Belenchia}}, \bibinfo {author} {\bibfnamefont {R.~M.}\ \bibnamefont {Wald}},
  \bibinfo {author} {\bibfnamefont {F.}~\bibnamefont {Giacomini}}, \bibinfo
  {author} {\bibfnamefont {E.}~\bibnamefont {Castro-Ruiz}}, \bibinfo {author}
  {\bibfnamefont {v.}~\bibnamefont {Brukner}},\ and\ \bibinfo {author}
  {\bibfnamefont {M.}~\bibnamefont {Aspelmeyer}},\ }\bibfield  {title}
  {\bibinfo {title} {{Quantum Superposition of Massive Objects and the
  Quantization of Gravity}},\ }\href
  {https://doi.org/10.1103/PhysRevD.98.126009} {\bibfield  {journal} {\bibinfo
  {journal} {Phys. Rev. D}\ }\textbf {\bibinfo {volume} {98}},\ \bibinfo
  {pages} {126009} (\bibinfo {year} {2018})},\ \Eprint
  {https://arxiv.org/abs/1807.07015} {arXiv:1807.07015 [quant-ph]} \BibitemShut
  {NoStop}%
\bibitem [{\citenamefont {Pontin}\ \emph {et~al.}(2018)\citenamefont {Pontin},
  \citenamefont {Bonaldi}, \citenamefont {Borrielli}, \citenamefont {Marconi},
  \citenamefont {Marino}, \citenamefont {Pandraud}, \citenamefont {Prodi},
  \citenamefont {Sarro}, \citenamefont {Serra},\ and\ \citenamefont
  {Marin}}]{Pontin:2016nem}%
  \BibitemOpen
  \bibfield  {author} {\bibinfo {author} {\bibfnamefont {A.}~\bibnamefont
  {Pontin}}, \bibinfo {author} {\bibfnamefont {M.}~\bibnamefont {Bonaldi}},
  \bibinfo {author} {\bibfnamefont {A.}~\bibnamefont {Borrielli}}, \bibinfo
  {author} {\bibfnamefont {L.}~\bibnamefont {Marconi}}, \bibinfo {author}
  {\bibfnamefont {F.}~\bibnamefont {Marino}}, \bibinfo {author} {\bibfnamefont
  {G.}~\bibnamefont {Pandraud}}, \bibinfo {author} {\bibfnamefont {G.~A.}\
  \bibnamefont {Prodi}}, \bibinfo {author} {\bibfnamefont {P.~M.}\ \bibnamefont
  {Sarro}}, \bibinfo {author} {\bibfnamefont {E.}~\bibnamefont {Serra}},\ and\
  \bibinfo {author} {\bibfnamefont {F.}~\bibnamefont {Marin}},\ }\bibfield
  {title} {\bibinfo {title} {{Quantum nondemolition measurement of optical
  field fluctuations by optomechanical interaction}},\ }\href
  {https://doi.org/10.1103/PhysRevA.97.033833} {\bibfield  {journal} {\bibinfo
  {journal} {Phys. Rev. A}\ }\textbf {\bibinfo {volume} {97}},\ \bibinfo
  {pages} {033833} (\bibinfo {year} {2018})},\ \Eprint
  {https://arxiv.org/abs/1611.00917} {arXiv:1611.00917 [quant-ph]} \BibitemShut
  {NoStop}%
\bibitem [{\citenamefont {Matsumoto}\ \emph {et~al.}(2015)\citenamefont
  {Matsumoto}, \citenamefont {Komori}, \citenamefont {Michimura}, \citenamefont
  {Hayase}, \citenamefont {Aso},\ and\ \citenamefont
  {Tsubono}}]{Matsumoto:2013sua}%
  \BibitemOpen
  \bibfield  {author} {\bibinfo {author} {\bibfnamefont {N.}~\bibnamefont
  {Matsumoto}}, \bibinfo {author} {\bibfnamefont {K.}~\bibnamefont {Komori}},
  \bibinfo {author} {\bibfnamefont {Y.}~\bibnamefont {Michimura}}, \bibinfo
  {author} {\bibfnamefont {G.}~\bibnamefont {Hayase}}, \bibinfo {author}
  {\bibfnamefont {Y.}~\bibnamefont {Aso}},\ and\ \bibinfo {author}
  {\bibfnamefont {K.}~\bibnamefont {Tsubono}},\ }\bibfield  {title} {\bibinfo
  {title} {{Classical Pendulum Feels Quantum Back-Action}},\ }\href
  {https://doi.org/10.1103/PhysRevA.92.033825} {\bibfield  {journal} {\bibinfo
  {journal} {Phys. Rev. A}\ }\textbf {\bibinfo {volume} {92}},\ \bibinfo
  {pages} {033825} (\bibinfo {year} {2015})},\ \Eprint
  {https://arxiv.org/abs/1312.5031} {arXiv:1312.5031 [quant-ph]} \BibitemShut
  {NoStop}%
\bibitem [{\citenamefont {Matsumoto}\ \emph {et~al.}(2014)\citenamefont
  {Matsumoto}, \citenamefont {Michimura}, \citenamefont {Aso},\ and\
  \citenamefont {Tsubono}}]{Matsumoto:2014fda}%
  \BibitemOpen
  \bibfield  {author} {\bibinfo {author} {\bibfnamefont {N.}~\bibnamefont
  {Matsumoto}}, \bibinfo {author} {\bibfnamefont {Y.}~\bibnamefont
  {Michimura}}, \bibinfo {author} {\bibfnamefont {Y.}~\bibnamefont {Aso}},\
  and\ \bibinfo {author} {\bibfnamefont {K.}~\bibnamefont {Tsubono}},\
  }\bibfield  {title} {\bibinfo {title} {{An optically trapped mirror for
  reaching the standard quantum limit}},\ }\href
  {https://doi.org/10.1364/OE.22.012915} {\bibfield  {journal} {\bibinfo
  {journal} {Opt. Express}\ }\textbf {\bibinfo {volume} {22}},\ \bibinfo
  {pages} {2915} (\bibinfo {year} {2014})},\ \Eprint
  {https://arxiv.org/abs/1405.4906} {arXiv:1405.4906 [physics.optics]}
  \BibitemShut {NoStop}%
\bibitem [{\citenamefont {Komori}\ \emph {et~al.}(2020)\citenamefont {Komori},
  \citenamefont {Enomoto}, \citenamefont {Ooi}, \citenamefont {Miyazaki},
  \citenamefont {Matsumoto}, \citenamefont {Sudhir}, \citenamefont
  {Michimura},\ and\ \citenamefont {Ando}}]{Komori:2019zlg}%
  \BibitemOpen
  \bibfield  {author} {\bibinfo {author} {\bibfnamefont {K.}~\bibnamefont
  {Komori}}, \bibinfo {author} {\bibfnamefont {Y.}~\bibnamefont {Enomoto}},
  \bibinfo {author} {\bibfnamefont {C.~P.}\ \bibnamefont {Ooi}}, \bibinfo
  {author} {\bibfnamefont {Y.}~\bibnamefont {Miyazaki}}, \bibinfo {author}
  {\bibfnamefont {N.}~\bibnamefont {Matsumoto}}, \bibinfo {author}
  {\bibfnamefont {V.}~\bibnamefont {Sudhir}}, \bibinfo {author} {\bibfnamefont
  {Y.}~\bibnamefont {Michimura}},\ and\ \bibinfo {author} {\bibfnamefont
  {M.}~\bibnamefont {Ando}},\ }\bibfield  {title} {\bibinfo {title}
  {{Attonewton-meter torque sensing with a macroscopic optomechanical torsion
  pendulum}},\ }\href {https://doi.org/10.1103/PhysRevA.101.011802} {\bibfield
  {journal} {\bibinfo  {journal} {Phys. Rev. A}\ }\textbf {\bibinfo {volume}
  {101}},\ \bibinfo {pages} {011802(R)} (\bibinfo {year} {2020})},\ \Eprint
  {https://arxiv.org/abs/1907.13139} {arXiv:1907.13139 [quant-ph]} \BibitemShut
  {NoStop}%
\bibitem [{\citenamefont {Sakata}\ \emph {et~al.}(2010)\citenamefont {Sakata},
  \citenamefont {Miyakawa}, \citenamefont {Nishizawa}, \citenamefont
  {Ishizaki},\ and\ \citenamefont {Kawamura}}]{Sakata:2010zz}%
  \BibitemOpen
  \bibfield  {author} {\bibinfo {author} {\bibfnamefont {S.}~\bibnamefont
  {Sakata}}, \bibinfo {author} {\bibfnamefont {O.}~\bibnamefont {Miyakawa}},
  \bibinfo {author} {\bibfnamefont {A.}~\bibnamefont {Nishizawa}}, \bibinfo
  {author} {\bibfnamefont {H.}~\bibnamefont {Ishizaki}},\ and\ \bibinfo
  {author} {\bibfnamefont {S.}~\bibnamefont {Kawamura}},\ }\bibfield  {title}
  {\bibinfo {title} {{Measurement of angular antispring effect in optical
  cavity by radiation pressure}},\ }\href
  {https://doi.org/10.1103/PhysRevD.81.064023} {\bibfield  {journal} {\bibinfo
  {journal} {Phys. Rev. D}\ }\textbf {\bibinfo {volume} {81}},\ \bibinfo
  {pages} {064023} (\bibinfo {year} {2010})}\BibitemShut {NoStop}%
\bibitem [{\citenamefont {Matsumoto}\ \emph {et~al.}(2019)\citenamefont
  {Matsumoto}, \citenamefont {Cata\~no Lopez}, \citenamefont {Sugawara},
  \citenamefont {Suzuki}, \citenamefont {Abe}, \citenamefont {Komori},
  \citenamefont {Michimura}, \citenamefont {Aso},\ and\ \citenamefont
  {Edamatsu}}]{Matsumoto:2018via}%
  \BibitemOpen
  \bibfield  {author} {\bibinfo {author} {\bibfnamefont {N.}~\bibnamefont
  {Matsumoto}}, \bibinfo {author} {\bibfnamefont {S.~B.}\ \bibnamefont
  {Cata\~no Lopez}}, \bibinfo {author} {\bibfnamefont {M.}~\bibnamefont
  {Sugawara}}, \bibinfo {author} {\bibfnamefont {S.}~\bibnamefont {Suzuki}},
  \bibinfo {author} {\bibfnamefont {N.}~\bibnamefont {Abe}}, \bibinfo {author}
  {\bibfnamefont {K.}~\bibnamefont {Komori}}, \bibinfo {author} {\bibfnamefont
  {Y.}~\bibnamefont {Michimura}}, \bibinfo {author} {\bibfnamefont
  {Y.}~\bibnamefont {Aso}},\ and\ \bibinfo {author} {\bibfnamefont
  {K.}~\bibnamefont {Edamatsu}},\ }\bibfield  {title} {\bibinfo {title}
  {{Demonstration of displacement sensing of a mg-scale pendulum for mm- and
  mg- scale gravity measurements}},\ }\href
  {https://doi.org/10.1103/PhysRevLett.122.071101} {\bibfield  {journal}
  {\bibinfo  {journal} {Phys. Rev. Lett.}\ }\textbf {\bibinfo {volume} {122}},\
  \bibinfo {pages} {071101} (\bibinfo {year} {2019})},\ \Eprint
  {https://arxiv.org/abs/1809.05081} {arXiv:1809.05081 [quant-ph]} \BibitemShut
  {NoStop}%
\bibitem [{\citenamefont {Corbitt}\ \emph {et~al.}(2007)\citenamefont
  {Corbitt}, \citenamefont {Wipf}, \citenamefont {Bodiya}, \citenamefont
  {Ottaway}, \citenamefont {Sigg}, \citenamefont {Smith}, \citenamefont
  {Whitcomb},\ and\ \citenamefont {Mavalvala}}]{Corbitt:2007spn}%
  \BibitemOpen
  \bibfield  {author} {\bibinfo {author} {\bibfnamefont {T.}~\bibnamefont
  {Corbitt}}, \bibinfo {author} {\bibfnamefont {C.}~\bibnamefont {Wipf}},
  \bibinfo {author} {\bibfnamefont {T.}~\bibnamefont {Bodiya}}, \bibinfo
  {author} {\bibfnamefont {D.}~\bibnamefont {Ottaway}}, \bibinfo {author}
  {\bibfnamefont {D.}~\bibnamefont {Sigg}}, \bibinfo {author} {\bibfnamefont
  {N.}~\bibnamefont {Smith}}, \bibinfo {author} {\bibfnamefont
  {S.}~\bibnamefont {Whitcomb}},\ and\ \bibinfo {author} {\bibfnamefont
  {N.}~\bibnamefont {Mavalvala}},\ }\bibfield  {title} {\bibinfo {title}
  {{Optical Dilution and Feedback Cooling of a Gram-Scale Oscillator to
  6.9~mK}},\ }\href {https://doi.org/10.1103/physrevlett.99.160801} {\bibfield
  {journal} {\bibinfo  {journal} {Phys. Rev. Lett.}\ }\textbf {\bibinfo
  {volume} {99}},\ \bibinfo {pages} {160801} (\bibinfo {year}
  {2007})}\BibitemShut {NoStop}%
\bibitem [{\citenamefont {Neben}\ \emph {et~al.}(2012)\citenamefont {Neben},
  \citenamefont {Bodiya}, \citenamefont {Wipf}, \citenamefont {Oelker},
  \citenamefont {Corbitt},\ and\ \citenamefont {Mavalvala}}]{Neben:2012edt}%
  \BibitemOpen
  \bibfield  {author} {\bibinfo {author} {\bibfnamefont {A.~R.}\ \bibnamefont
  {Neben}}, \bibinfo {author} {\bibfnamefont {T.~P.}\ \bibnamefont {Bodiya}},
  \bibinfo {author} {\bibfnamefont {C.}~\bibnamefont {Wipf}}, \bibinfo {author}
  {\bibfnamefont {E.}~\bibnamefont {Oelker}}, \bibinfo {author} {\bibfnamefont
  {T.}~\bibnamefont {Corbitt}},\ and\ \bibinfo {author} {\bibfnamefont
  {N.}~\bibnamefont {Mavalvala}},\ }\bibfield  {title} {\bibinfo {title}
  {{Structural thermal noise in gram-scale mirror oscillators}},\ }\bibfield
  {journal} {\bibinfo  {journal} {New J. Phys.}\ }\textbf {\bibinfo {volume}
  {14}},\ \href {https://doi.org/10.1088/1367-2630/14/11/115008}
  {10.1088/1367-2630/14/11/115008} (\bibinfo {year} {2012})\BibitemShut
  {NoStop}%
\bibitem [{\citenamefont {Abbott}\ \emph {et~al.}(2009)\citenamefont {Abbott}
  \emph {et~al.}}]{LIGOScientific:2009mif}%
  \BibitemOpen
  \bibfield  {author} {\bibinfo {author} {\bibfnamefont {B.}~\bibnamefont
  {Abbott}} \emph {et~al.} (\bibinfo {collaboration} {LIGO Scientific}),\
  }\bibfield  {title} {\bibinfo {title} {{Observation of a kilogram-scale
  oscillator near its quantum ground state}},\ }\href
  {https://doi.org/10.1088/1367-2630/11/7/073032} {\bibfield  {journal}
  {\bibinfo  {journal} {New J. Phys.}\ }\textbf {\bibinfo {volume} {11}},\
  \bibinfo {pages} {073032} (\bibinfo {year} {2009})}\BibitemShut {NoStop}%
\bibitem [{\citenamefont {Yu}\ \emph {et~al.}(2020)\citenamefont {Yu} \emph
  {et~al.}}]{LIGOScientific:2020luc}%
  \BibitemOpen
  \bibfield  {author} {\bibinfo {author} {\bibfnamefont {H.}~\bibnamefont {Yu}}
  \emph {et~al.} (\bibinfo {collaboration} {LIGO Scientific}),\ }\bibfield
  {title} {\bibinfo {title} {{Quantum correlations between light and the
  kilogram-mass mirrors of LIGO}},\ }\href
  {https://doi.org/10.1038/s41586-020-2420-8} {\bibfield  {journal} {\bibinfo
  {journal} {Nature}\ }\textbf {\bibinfo {volume} {583}},\ \bibinfo {pages}
  {43} (\bibinfo {year} {2020})},\ \Eprint {https://arxiv.org/abs/2002.01519}
  {arXiv:2002.01519 [quant-ph]} \BibitemShut {NoStop}%
\bibitem [{\citenamefont {Chan}\ \emph {et~al.}(2011)\citenamefont {Chan},
  \citenamefont {Alegre}, \citenamefont {Safavi-Naeini}, \citenamefont {Hill},
  \citenamefont {Krause}, \citenamefont {Gr\"oblacher}, \citenamefont
  {Aspelmeyer},\ and\ \citenamefont {Painter}}]{Chan:2011ivv}%
  \BibitemOpen
  \bibfield  {author} {\bibinfo {author} {\bibfnamefont {J.}~\bibnamefont
  {Chan}}, \bibinfo {author} {\bibfnamefont {T.~P.~M.}\ \bibnamefont {Alegre}},
  \bibinfo {author} {\bibfnamefont {A.~H.}\ \bibnamefont {Safavi-Naeini}},
  \bibinfo {author} {\bibfnamefont {J.~T.}\ \bibnamefont {Hill}}, \bibinfo
  {author} {\bibfnamefont {A.}~\bibnamefont {Krause}}, \bibinfo {author}
  {\bibfnamefont {S.}~\bibnamefont {Gr\"oblacher}}, \bibinfo {author}
  {\bibfnamefont {M.}~\bibnamefont {Aspelmeyer}},\ and\ \bibinfo {author}
  {\bibfnamefont {O.}~\bibnamefont {Painter}},\ }\bibfield  {title} {\bibinfo
  {title} {{Laser cooling of a nanomechanical oscillator into its quantum
  ground state}},\ }\href {https://doi.org/10.1038/nature10461} {\bibfield
  {journal} {\bibinfo  {journal} {Nature}\ }\textbf {\bibinfo {volume} {478}},\
  \bibinfo {pages} {89} (\bibinfo {year} {2011})}\BibitemShut {NoStop}%
\bibitem [{\citenamefont {Teufel}\ \emph {et~al.}(2011)\citenamefont {Teufel},
  \citenamefont {Donner}, \citenamefont {Li}, \citenamefont {Harlow},
  \citenamefont {Allman}, \citenamefont {Cicak}, \citenamefont {Sirois},
  \citenamefont {Whittaker}, \citenamefont {Lehnert},\ and\ \citenamefont
  {Simmonds}}]{Teufel:2011smx}%
  \BibitemOpen
  \bibfield  {author} {\bibinfo {author} {\bibfnamefont {J.~D.}\ \bibnamefont
  {Teufel}}, \bibinfo {author} {\bibfnamefont {T.}~\bibnamefont {Donner}},
  \bibinfo {author} {\bibfnamefont {D.}~\bibnamefont {Li}}, \bibinfo {author}
  {\bibfnamefont {J.~W.}\ \bibnamefont {Harlow}}, \bibinfo {author}
  {\bibfnamefont {M.~S.}\ \bibnamefont {Allman}}, \bibinfo {author}
  {\bibfnamefont {K.}~\bibnamefont {Cicak}}, \bibinfo {author} {\bibfnamefont
  {A.~J.}\ \bibnamefont {Sirois}}, \bibinfo {author} {\bibfnamefont {J.~D.}\
  \bibnamefont {Whittaker}}, \bibinfo {author} {\bibfnamefont {K.~W.}\
  \bibnamefont {Lehnert}},\ and\ \bibinfo {author} {\bibfnamefont {R.~W.}\
  \bibnamefont {Simmonds}},\ }\bibfield  {title} {\bibinfo {title} {{Sideband
  cooling of micromechanical motion to the quantum ground state}},\ }\href
  {https://doi.org/10.1038/nature10261} {\bibfield  {journal} {\bibinfo
  {journal} {Nature}\ }\textbf {\bibinfo {volume} {475}},\ \bibinfo {pages}
  {359} (\bibinfo {year} {2011})}\BibitemShut {NoStop}%
\bibitem [{\citenamefont {Peterson}\ \emph {et~al.}(2016)\citenamefont
  {Peterson}, \citenamefont {Purdy}, \citenamefont {Kampel}, \citenamefont
  {Andrews}, \citenamefont {Yu}, \citenamefont {Lehnert},\ and\ \citenamefont
  {Regal}}]{Peterson:2016ayo}%
  \BibitemOpen
  \bibfield  {author} {\bibinfo {author} {\bibfnamefont {R.~W.}\ \bibnamefont
  {Peterson}}, \bibinfo {author} {\bibfnamefont {T.~P.}\ \bibnamefont {Purdy}},
  \bibinfo {author} {\bibfnamefont {N.~S.}\ \bibnamefont {Kampel}}, \bibinfo
  {author} {\bibfnamefont {R.~W.}\ \bibnamefont {Andrews}}, \bibinfo {author}
  {\bibfnamefont {P.~L.}\ \bibnamefont {Yu}}, \bibinfo {author} {\bibfnamefont
  {K.~W.}\ \bibnamefont {Lehnert}},\ and\ \bibinfo {author} {\bibfnamefont
  {C.~A.}\ \bibnamefont {Regal}},\ }\bibfield  {title} {\bibinfo {title}
  {{Laser Cooling of a Micromechanical Membrane to the Quantum Backaction
  Limit}},\ }\href {https://doi.org/10.1103/physrevlett.116.063601} {\bibfield
  {journal} {\bibinfo  {journal} {Phys. Rev. Lett.}\ }\textbf {\bibinfo
  {volume} {116}},\ \bibinfo {pages} {063601} (\bibinfo {year}
  {2016})}\BibitemShut {NoStop}%
\bibitem [{\citenamefont {Graham}\ \emph {et~al.}(2016)\citenamefont {Graham},
  \citenamefont {Kaplan}, \citenamefont {Mardon}, \citenamefont {Rajendran},\
  and\ \citenamefont {Terrano}}]{Graham:2015ifn}%
  \BibitemOpen
  \bibfield  {author} {\bibinfo {author} {\bibfnamefont {P.~W.}\ \bibnamefont
  {Graham}}, \bibinfo {author} {\bibfnamefont {D.~E.}\ \bibnamefont {Kaplan}},
  \bibinfo {author} {\bibfnamefont {J.}~\bibnamefont {Mardon}}, \bibinfo
  {author} {\bibfnamefont {S.}~\bibnamefont {Rajendran}},\ and\ \bibinfo
  {author} {\bibfnamefont {W.~A.}\ \bibnamefont {Terrano}},\ }\bibfield
  {title} {\bibinfo {title} {{Dark Matter Direct Detection with
  Accelerometers}},\ }\href {https://doi.org/10.1103/PhysRevD.93.075029}
  {\bibfield  {journal} {\bibinfo  {journal} {Phys. Rev. D}\ }\textbf {\bibinfo
  {volume} {93}},\ \bibinfo {pages} {075029} (\bibinfo {year} {2016})},\
  \Eprint {https://arxiv.org/abs/1512.06165} {arXiv:1512.06165 [hep-ph]}
  \BibitemShut {NoStop}%
\bibitem [{\citenamefont {Pierce}\ \emph {et~al.}(2018)\citenamefont {Pierce},
  \citenamefont {Riles},\ and\ \citenamefont {Zhao}}]{Pierce:2018xmy}%
  \BibitemOpen
  \bibfield  {author} {\bibinfo {author} {\bibfnamefont {A.}~\bibnamefont
  {Pierce}}, \bibinfo {author} {\bibfnamefont {K.}~\bibnamefont {Riles}},\ and\
  \bibinfo {author} {\bibfnamefont {Y.}~\bibnamefont {Zhao}},\ }\bibfield
  {title} {\bibinfo {title} {{Searching for Dark Photon Dark Matter with
  Gravitational Wave Detectors}},\ }\href
  {https://doi.org/10.1103/PhysRevLett.121.061102} {\bibfield  {journal}
  {\bibinfo  {journal} {Phys. Rev. Lett.}\ }\textbf {\bibinfo {volume} {121}},\
  \bibinfo {pages} {061102} (\bibinfo {year} {2018})},\ \Eprint
  {https://arxiv.org/abs/1801.10161} {arXiv:1801.10161 [hep-ph]} \BibitemShut
  {NoStop}%
\bibitem [{\citenamefont {Carney}\ \emph {et~al.}(2021)\citenamefont {Carney},
  \citenamefont {Hook}, \citenamefont {Liu}, \citenamefont {Taylor},\ and\
  \citenamefont {Zhao}}]{Carney:2019cio}%
  \BibitemOpen
  \bibfield  {author} {\bibinfo {author} {\bibfnamefont {D.}~\bibnamefont
  {Carney}}, \bibinfo {author} {\bibfnamefont {A.}~\bibnamefont {Hook}},
  \bibinfo {author} {\bibfnamefont {Z.}~\bibnamefont {Liu}}, \bibinfo {author}
  {\bibfnamefont {J.~M.}\ \bibnamefont {Taylor}},\ and\ \bibinfo {author}
  {\bibfnamefont {Y.}~\bibnamefont {Zhao}},\ }\bibfield  {title} {\bibinfo
  {title} {{Ultralight dark matter detection with mechanical quantum
  sensors}},\ }\href {https://doi.org/10.1088/1367-2630/abd9e7} {\bibfield
  {journal} {\bibinfo  {journal} {New J. Phys.}\ }\textbf {\bibinfo {volume}
  {23}},\ \bibinfo {pages} {023041} (\bibinfo {year} {2021})},\ \Eprint
  {https://arxiv.org/abs/1908.04797} {arXiv:1908.04797 [hep-ph]} \BibitemShut
  {NoStop}%
\bibitem [{\citenamefont {Sidles}\ and\ \citenamefont
  {Sigg}(2006)}]{Sidles:2006vzf}%
  \BibitemOpen
  \bibfield  {author} {\bibinfo {author} {\bibfnamefont {J.~A.}\ \bibnamefont
  {Sidles}}\ and\ \bibinfo {author} {\bibfnamefont {D.}~\bibnamefont {Sigg}},\
  }\bibfield  {title} {\bibinfo {title} {{Optical torques in suspended
  Fabry\textendash{}Perot interferometers}},\ }\href
  {https://doi.org/10.1016/j.physleta.2006.01.051} {\bibfield  {journal}
  {\bibinfo  {journal} {Phys. Lett. A}\ }\textbf {\bibinfo {volume} {354}},\
  \bibinfo {pages} {167} (\bibinfo {year} {2006})}\BibitemShut {NoStop}%
\bibitem [{\citenamefont {Kelley}\ \emph {et~al.}(2015)\citenamefont {Kelley},
  \citenamefont {Lough}, \citenamefont {Manga\~na{-}Sandoval}, \citenamefont
  {Perreca},\ and\ \citenamefont {Ballmer}}]{Kelley:2015axa}%
  \BibitemOpen
  \bibfield  {author} {\bibinfo {author} {\bibfnamefont {D.}~\bibnamefont
  {Kelley}}, \bibinfo {author} {\bibfnamefont {J.}~\bibnamefont {Lough}},
  \bibinfo {author} {\bibfnamefont {F.}~\bibnamefont {Manga\~na{-}Sandoval}},
  \bibinfo {author} {\bibfnamefont {A.}~\bibnamefont {Perreca}},\ and\ \bibinfo
  {author} {\bibfnamefont {S.~W.}\ \bibnamefont {Ballmer}},\ }\bibfield
  {title} {\bibinfo {title} {{Observation of photo-thermal feed-back in a
  stable dual-carrier optical spring}},\ }\href
  {https://doi.org/10.1103/PhysRevD.92.062003} {\bibfield  {journal} {\bibinfo
  {journal} {Phys. Rev. D}\ }\textbf {\bibinfo {volume} {92}},\ \bibinfo
  {pages} {062003} (\bibinfo {year} {2015})},\ \Eprint
  {https://arxiv.org/abs/1507.02979} {arXiv:1507.02979 [physics.optics]}
  \BibitemShut {NoStop}%
\bibitem [{\citenamefont {Barsotti}\ \emph {et~al.}(2010)\citenamefont
  {Barsotti}, \citenamefont {Evans},\ and\ \citenamefont
  {Fritschel}}]{Barsotti:2010zz}%
  \BibitemOpen
  \bibfield  {author} {\bibinfo {author} {\bibfnamefont {L.}~\bibnamefont
  {Barsotti}}, \bibinfo {author} {\bibfnamefont {M.}~\bibnamefont {Evans}},\
  and\ \bibinfo {author} {\bibfnamefont {P.}~\bibnamefont {Fritschel}},\
  }\bibfield  {title} {\bibinfo {title} {{Alignment sensing and control in
  advanced LIGO}},\ }\href {https://doi.org/10.1088/0264-9381/27/8/084026}
  {\bibfield  {journal} {\bibinfo  {journal} {Class. Quant. Grav.}\ }\textbf
  {\bibinfo {volume} {27}},\ \bibinfo {pages} {084026} (\bibinfo {year}
  {2010})}\BibitemShut {NoStop}%
\bibitem [{\citenamefont {Liu}\ \emph {et~al.}(2018)\citenamefont {Liu},
  \citenamefont {Bossilkov}, \citenamefont {Blair}, \citenamefont {Zhao},
  \citenamefont {Ju},\ and\ \citenamefont {Blair}}]{Liu:2018dfs}%
  \BibitemOpen
  \bibfield  {author} {\bibinfo {author} {\bibfnamefont {J.}~\bibnamefont
  {Liu}}, \bibinfo {author} {\bibfnamefont {V.}~\bibnamefont {Bossilkov}},
  \bibinfo {author} {\bibfnamefont {C.}~\bibnamefont {Blair}}, \bibinfo
  {author} {\bibfnamefont {C.}~\bibnamefont {Zhao}}, \bibinfo {author}
  {\bibfnamefont {L.}~\bibnamefont {Ju}},\ and\ \bibinfo {author}
  {\bibfnamefont {D.}~\bibnamefont {Blair}},\ }\bibfield  {title} {\bibinfo
  {title} {{Angular instability in high optical power suspended cavities}},\
  }\href {https://doi.org/10.1063/1.5049508} {\bibfield  {journal} {\bibinfo
  {journal} {Rev. Sci. Instrum.}\ }\textbf {\bibinfo {volume} {89}},\ \bibinfo
  {pages} {124503} (\bibinfo {year} {2018})},\ \Eprint
  {https://arxiv.org/abs/1812.00813} {arXiv:1812.00813 [astro-ph.IM]}
  \BibitemShut {NoStop}%
\bibitem [{\citenamefont {Enomoto}\ \emph {et~al.}(2016)\citenamefont
  {Enomoto}, \citenamefont {Nagano}, \citenamefont {Nakano}, \citenamefont
  {Furusawa},\ and\ \citenamefont {Kawamura}}]{Enomoto:2016lee}%
  \BibitemOpen
  \bibfield  {author} {\bibinfo {author} {\bibfnamefont {Y.}~\bibnamefont
  {Enomoto}}, \bibinfo {author} {\bibfnamefont {K.}~\bibnamefont {Nagano}},
  \bibinfo {author} {\bibfnamefont {M.}~\bibnamefont {Nakano}}, \bibinfo
  {author} {\bibfnamefont {A.}~\bibnamefont {Furusawa}},\ and\ \bibinfo
  {author} {\bibfnamefont {S.}~\bibnamefont {Kawamura}},\ }\bibfield  {title}
  {\bibinfo {title} {{Observation of reduction of radiation-pressure-induced
  rotational anti-spring effect on a 23 mg mirror in a Fabry-Perot cavity}},\
  }\href {https://doi.org/10.1088/0264-9381/33/14/145002} {\bibfield  {journal}
  {\bibinfo  {journal} {Class. Quant. Grav.}\ }\textbf {\bibinfo {volume}
  {33}},\ \bibinfo {pages} {145002} (\bibinfo {year} {2016})}\BibitemShut
  {NoStop}%
\bibitem [{\citenamefont {Nagano}\ \emph {et~al.}(2016)\citenamefont {Nagano},
  \citenamefont {Enomoto}, \citenamefont {Nakano}, \citenamefont {Furusawa},\
  and\ \citenamefont {Kawamura}}]{Nagano:2016wcg}%
  \BibitemOpen
  \bibfield  {author} {\bibinfo {author} {\bibfnamefont {K.}~\bibnamefont
  {Nagano}}, \bibinfo {author} {\bibfnamefont {Y.}~\bibnamefont {Enomoto}},
  \bibinfo {author} {\bibfnamefont {M.}~\bibnamefont {Nakano}}, \bibinfo
  {author} {\bibfnamefont {A.}~\bibnamefont {Furusawa}},\ and\ \bibinfo
  {author} {\bibfnamefont {S.}~\bibnamefont {Kawamura}},\ }\bibfield  {title}
  {\bibinfo {title} {{Mitigation of radiation-pressure-induced angular
  instability of a Fabry-Perot cavity consisting of suspended mirrors}},\
  }\href {https://doi.org/10.1016/j.physleta.2016.09.056} {\bibfield  {journal}
  {\bibinfo  {journal} {Phys. Lett. A}\ }\textbf {\bibinfo {volume} {380}},\
  \bibinfo {pages} {3871} (\bibinfo {year} {2016})}\BibitemShut {NoStop}%
\bibitem [{\citenamefont {Michimura}\ and\ \citenamefont
  {Komori}(2020)}]{Michimura:2020yvn}%
  \BibitemOpen
  \bibfield  {author} {\bibinfo {author} {\bibfnamefont {Y.}~\bibnamefont
  {Michimura}}\ and\ \bibinfo {author} {\bibfnamefont {K.}~\bibnamefont
  {Komori}},\ }\bibfield  {title} {\bibinfo {title} {{Quantum sensing with
  milligram scale optomechanical systems}},\ }\href
  {https://doi.org/10.1140/epjd/e2020-10185-5} {\bibfield  {journal} {\bibinfo
  {journal} {Eur. Phys. J. D}\ }\textbf {\bibinfo {volume} {74}},\ \bibinfo
  {pages} {126} (\bibinfo {year} {2020})},\ \Eprint
  {https://arxiv.org/abs/2003.13906} {arXiv:2003.13906 [quant-ph]} \BibitemShut
  {NoStop}%
\bibitem [{\citenamefont {Levin}(1998)}]{Levin:1997kv}%
  \BibitemOpen
  \bibfield  {author} {\bibinfo {author} {\bibfnamefont {Y.}~\bibnamefont
  {Levin}},\ }\bibfield  {title} {\bibinfo {title} {{Internal thermal noise in
  the LIGO test masses: A Direct approach}},\ }\href
  {https://doi.org/10.1103/PhysRevD.57.659} {\bibfield  {journal} {\bibinfo
  {journal} {Phys. Rev. D}\ }\textbf {\bibinfo {volume} {57}},\ \bibinfo
  {pages} {659} (\bibinfo {year} {1998})},\ \Eprint
  {https://arxiv.org/abs/gr-qc/9707013} {arXiv:gr-qc/9707013} \BibitemShut
  {NoStop}%
\bibitem [{\citenamefont {Bondu}\ \emph {et~al.}(1998)\citenamefont {Bondu},
  \citenamefont {Hello},\ and\ \citenamefont {Vinet}}]{Bondu:1998sm}%
  \BibitemOpen
  \bibfield  {author} {\bibinfo {author} {\bibfnamefont {F.}~\bibnamefont
  {Bondu}}, \bibinfo {author} {\bibfnamefont {P.}~\bibnamefont {Hello}},\ and\
  \bibinfo {author} {\bibfnamefont {J.~Y.}\ \bibnamefont {Vinet}},\ }\bibfield
  {title} {\bibinfo {title} {{Thermal noise in mirrors of interferometric
  gravitational wave antennas}},\ }\href
  {https://doi.org/10.1016/S0375-9601(98)00450-2} {\bibfield  {journal}
  {\bibinfo  {journal} {Phys. Lett. A}\ }\textbf {\bibinfo {volume} {246}},\
  \bibinfo {pages} {227} (\bibinfo {year} {1998})}\BibitemShut {NoStop}%
\bibitem [{\citenamefont {Villar}\ \emph {et~al.}(2010)\citenamefont {Villar},
  \citenamefont {Black}, \citenamefont {DeSalvo}, \citenamefont {Libbrecht},
  \citenamefont {Michel}, \citenamefont {Morgado}, \citenamefont {Pinard},
  \citenamefont {Pinto}, \citenamefont {Pierro}, \citenamefont {Galdi},
  \citenamefont {Principe},\ and\ \citenamefont {Taurasi}}]{Villar:2010kf}%
  \BibitemOpen
  \bibfield  {author} {\bibinfo {author} {\bibfnamefont {A.~E.}\ \bibnamefont
  {Villar}}, \bibinfo {author} {\bibfnamefont {E.~D.}\ \bibnamefont {Black}},
  \bibinfo {author} {\bibfnamefont {R.}~\bibnamefont {DeSalvo}}, \bibinfo
  {author} {\bibfnamefont {K.~G.}\ \bibnamefont {Libbrecht}}, \bibinfo {author}
  {\bibfnamefont {C.}~\bibnamefont {Michel}}, \bibinfo {author} {\bibfnamefont
  {N.}~\bibnamefont {Morgado}}, \bibinfo {author} {\bibfnamefont
  {L.}~\bibnamefont {Pinard}}, \bibinfo {author} {\bibfnamefont {I.~M.}\
  \bibnamefont {Pinto}}, \bibinfo {author} {\bibfnamefont {V.}~\bibnamefont
  {Pierro}}, \bibinfo {author} {\bibfnamefont {V.}~\bibnamefont {Galdi}},
  \bibinfo {author} {\bibfnamefont {M.}~\bibnamefont {Principe}},\ and\
  \bibinfo {author} {\bibfnamefont {I.}~\bibnamefont {Taurasi}},\ }\bibfield
  {title} {\bibinfo {title} {{Measurement of Thermal Noise in Multilayer
  Coatings with Optimized Layer Thickness}},\ }\href
  {https://doi.org/10.1103/PhysRevD.81.122001} {\bibfield  {journal} {\bibinfo
  {journal} {Phys. Rev. D}\ }\textbf {\bibinfo {volume} {81}},\ \bibinfo
  {pages} {122001} (\bibinfo {year} {2010})},\ \Eprint
  {https://arxiv.org/abs/1004.1223} {arXiv:1004.1223 [gr-qc]} \BibitemShut
  {NoStop}%
\bibitem [{\citenamefont {Shoemaker}\ \emph {et~al.}(1988)\citenamefont
  {Shoemaker}, \citenamefont {Schilling}, \citenamefont {Schnupp},
  \citenamefont {Winkler}, \citenamefont {Maischberger},\ and\ \citenamefont
  {Rudiger}}]{Shoemaker:1987ft}%
  \BibitemOpen
  \bibfield  {author} {\bibinfo {author} {\bibfnamefont {D.}~\bibnamefont
  {Shoemaker}}, \bibinfo {author} {\bibfnamefont {R.}~\bibnamefont
  {Schilling}}, \bibinfo {author} {\bibfnamefont {L.}~\bibnamefont {Schnupp}},
  \bibinfo {author} {\bibfnamefont {W.}~\bibnamefont {Winkler}}, \bibinfo
  {author} {\bibfnamefont {K.}~\bibnamefont {Maischberger}},\ and\ \bibinfo
  {author} {\bibfnamefont {A.}~\bibnamefont {Rudiger}},\ }\bibfield  {title}
  {\bibinfo {title} {{Noise Behavior of the Garching 30-meter Prototype
  Gravitational Wave Detector}},\ }\href
  {https://doi.org/10.1103/PhysRevD.38.423} {\bibfield  {journal} {\bibinfo
  {journal} {Phys. Rev. D}\ }\textbf {\bibinfo {volume} {38}},\ \bibinfo
  {pages} {423} (\bibinfo {year} {1988})}\BibitemShut {NoStop}%
\bibitem [{\citenamefont {Saulson}(1990)}]{Saulson:1990jc}%
  \BibitemOpen
  \bibfield  {author} {\bibinfo {author} {\bibfnamefont {P.~R.}\ \bibnamefont
  {Saulson}},\ }\bibfield  {title} {\bibinfo {title} {{Thermal noise in
  mechanical experiments}},\ }\href {https://doi.org/10.1103/PhysRevD.42.2437}
  {\bibfield  {journal} {\bibinfo  {journal} {Phys. Rev. D}\ }\textbf {\bibinfo
  {volume} {42}},\ \bibinfo {pages} {2437} (\bibinfo {year}
  {1990})}\BibitemShut {NoStop}%
\bibitem [{\citenamefont {Bondu}\ \emph {et~al.}(1996)\citenamefont {Bondu},
  \citenamefont {Fritschel}, \citenamefont {Man},\ and\ \citenamefont
  {Brillet}}]{Bondu:1996rnn}%
  \BibitemOpen
  \bibfield  {author} {\bibinfo {author} {\bibfnamefont {F.}~\bibnamefont
  {Bondu}}, \bibinfo {author} {\bibfnamefont {P.}~\bibnamefont {Fritschel}},
  \bibinfo {author} {\bibfnamefont {C.~N.}\ \bibnamefont {Man}},\ and\ \bibinfo
  {author} {\bibfnamefont {A.}~\bibnamefont {Brillet}},\ }\bibfield  {title}
  {\bibinfo {title} {{Ultrahigh-spectral-purity laser for the VIRGO
  experiment}},\ }\href {https://doi.org/10.1364/OL.21.000582} {\bibfield
  {journal} {\bibinfo  {journal} {Opt. Lett.}\ }\textbf {\bibinfo {volume}
  {21}},\ \bibinfo {pages} {582} (\bibinfo {year} {1996})}\BibitemShut
  {NoStop}%
\bibitem [{\citenamefont {Conti}\ \emph {et~al.}(2003)\citenamefont {Conti},
  \citenamefont {Rosa},\ and\ \citenamefont {Marin}}]{Conti:2003}%
  \BibitemOpen
  \bibfield  {author} {\bibinfo {author} {\bibfnamefont {L.}~\bibnamefont
  {Conti}}, \bibinfo {author} {\bibfnamefont {M.~D.}\ \bibnamefont {Rosa}},\
  and\ \bibinfo {author} {\bibfnamefont {F.}~\bibnamefont {Marin}},\ }\bibfield
   {title} {\bibinfo {title} {High-spectral-purity laser system for the auriga
  detector optical readout},\ }\href {https://doi.org/10.1364/JOSAB.20.000462}
  {\bibfield  {journal} {\bibinfo  {journal} {J. Opt. Soc. Am. B}\ }\textbf
  {\bibinfo {volume} {20}},\ \bibinfo {pages} {462} (\bibinfo {year}
  {2003})}\BibitemShut {NoStop}%
\bibitem [{\citenamefont {Conti}\ \emph {et~al.}(2000)\citenamefont {Conti},
  \citenamefont {De~Rosa},\ and\ \citenamefont {Marin}}]{Conti:2000nhd}%
  \BibitemOpen
  \bibfield  {author} {\bibinfo {author} {\bibfnamefont {L.}~\bibnamefont
  {Conti}}, \bibinfo {author} {\bibfnamefont {M.}~\bibnamefont {De~Rosa}},\
  and\ \bibinfo {author} {\bibfnamefont {F.}~\bibnamefont {Marin}},\ }\bibfield
   {title} {\bibinfo {title} {{Low-amplitude-noise laser for AURIGA detector
  optical readout}},\ }\href {https://doi.org/10.1364/AO.39.005732} {\bibfield
  {journal} {\bibinfo  {journal} {Appl. Opt.}\ }\textbf {\bibinfo {volume}
  {39}},\ \bibinfo {pages} {5732} (\bibinfo {year} {2000})}\BibitemShut
  {NoStop}%
\end{thebibliography}%

\end{document}